\documentclass[acmsmall]{acmart}
\AtBeginDocument{%
  \providecommand\BibTeX{{%
    \normalfont B\kern-0.5em{\scshape i\kern-0.25em b}\kern-0.8em\TeX}}}

\setcopyright{acmcopyright}
\copyrightyear{2021}
\acmYear{2021}
\acmDOI{10.1145/1122445.1122456}

\acmJournal{TCPS}
\acmVolume{1}
\acmNumber{1}
\acmArticle{1}
\acmMonth{1}

\usepackage{tabularx}
\usepackage[mode=text]{siunitx}
\usepackage{multicol}

\begin{document}

\title{Edge computing for cyber-physical systems}
\subtitle{A systematic mapping study emphasizing trustworthiness}

\author{José Manuel Gaspar Sánchez}
\authornote{Both authors contributed equally to this research.}
\email{jmgs@kth.se}
\author{Nils J\"orgensen}
\authornotemark[1]
\email{nilsjor@kth.se}

\author{Martin Törngren}
\email{martint@kth.se}

\affiliation{%
  \institution{KTH Royal Institute of Technology}
  \streetaddress{Brinellvägen 86}
  \city{Stockholm}
  \country{Sweden}
  \postcode{114 28}
}

\author{Rafia Inam}
\email{rafia.inam@ericsson.com}

\affiliation{%
  \institution{Ericsson Research}
  \country{Sweden}
}

\author{Andrii Berezovskyi}
\email{andriib@kth.se}
\author{Lei Feng}
\email{lfeng@kth.se}
\author{Elena Fersman}
\email{fersman@kth.se}
\author{Muhammad Rusyadi Ramli}
\email{ramli2@kth.se}
\author{Kaige Tan}
\email{kaiget@kth.se}

\affiliation{%
  \institution{KTH Royal Institute of Technology}
  \streetaddress{Brinellvägen 86}
  \city{Stockholm}
  \country{Sweden}
  \postcode{114 28}
}

\renewcommand{\shortauthors}{Gaspar Sánchez and Jörgensen, et al.}

\begin{abstract}
    Edge computing is projected to have profound implications in the coming decades, proposed to provide solutions for applications such as augmented reality, predictive functionalities, and collaborative Cyber-Physical Systems (CPS). For such applications, edge computing addresses the new computational needs, as well as privacy, availability, and real-time constraints, by providing local high-performance computing capabilities to deal with the limitations and constraints of cloud and embedded systems. 
    Edge computing is today driven by strong market forces stemming from IT/cloud, telecom, and networking - with corresponding multiple interpretations of "edge computing" (e.g. device edge, network edge, distributed cloud, etc.). Considering the strong drivers for edge-computing and the relative novelty of the field, it becomes important to understand the specific requirements and characteristics of edge-based CPS, and to ensure that research is guided adequately, e.g. avoiding specific gaps.
    
    Our interests lie in the applications of edge computing as part of CPS, where several properties (or attributes) of trustworthiness, including safety, security, and predictability/availability are of particular concern, each facing challenges for the introduction of edge-based CPS. We present the results of a systematic mapping study, a kind of systematic literature survey, investigating the use of edge computing for CPS with a special emphasis on trustworthiness. The main contributions of this study are a detailed description of the current research efforts in edge-based CPS and the identification and discussion of trends and research gaps. 
    The results show that the main body of research in edge-based CPS only to a very limited extent consider key attributes of system trustworthiness, despite many efforts referring to critical CPS and applications like intelligent transportation. More research and industrial efforts will be needed on aspects of trustworthiness of future edge-based CPS including their experimental evaluation. Such research needs to consider the multiple interrelated attributes of trustworthiness including safety, security, and predictability, and new methodologies and architectures to address them. It is further important to provide bridges and collaboration between edge computing and CPS disciplines.
\end{abstract}

\ccsdesc[500]{General and reference~Surveys and overviews}
\ccsdesc[500]{Computer systems organization~Embedded and cyber-physical systems}
\ccsdesc[300]{Computer systems organization~Dependable and fault-tolerant systems and networks}

\keywords{edge computing, fog computing, mobile edge computing, cloudlet, cyber-physical systems, trustworthiness, safety, security, predictability, dependability, critical systems}

\maketitle

\section{Introduction}   



Adding to the landscape of embedded systems, cloud computing, networking, and telecommunications -- \textit{edge computing} is proposed to provide solutions for various cyber-physical systems (CPS) and applications, such as augmented reality, predictive functions e.g. for anomaly detection, and collaborative CPS to name a few. These applications often share requirements on high availability, real-time behavior, domain-specific sensitive data, while more and more involving huge amounts of data and corresponding processing demands. 

A key advantage of edge computing is localized and enhanced computational performance, which reduces costs on the device/embedded systems side because of less computing and storage resources and overcomes the shortcoming on latency, bandwidth, and privacy issues of the centralized cloud-based solutions. 
By adding a new third tier of computing to address these requirements and limitations, edge computing is projected to have profound implications in the coming decades, 
~\cite{sat17, Ahmed2017, 8100873, Khan2019, toerngren16}. 


As a consequence, edge computing is today driven by strong market forces stemming from IT/cloud, telecom, and networking - with corresponding multiple interpretations of "edge computing", including in terms of where the edge lies, e.g. device edge, network edge, distributed cloud, etc. Such interpretations include

\begin {itemize}
\item \textit{multi-access edge computing} (MEC) -- a term coined by the  European Telecommunications Standards Institute\footnote{\url{https://www.etsi.org/technologies/multi-access-edge-computing} (accessed Dec. 21, 2020)}, previously referred to as \textit{mobile} edge computing. This edge computing concept is closely associated with telecom and 5G networks, for example exploiting base stations as compute facilities ~\cite{Abbas.2018},
\item \textit{fog computing} -- as an extension of cloud computing that beyond the cloud leverages additional localized resources such as routers and gateways~\cite{Bonomi.2012},
\item \textit{cloudlets} -- as clusters of trusted computers with a strong connection to the Internet that is utilized by nearby mobile devices~\cite{Satyanarayanan.2009}. 
\end {itemize}



The focus of this study lies the in the intersection of the various {edge computing} paradigms and {cyber-physical systems} and applications.
CPS represent the “Integration of computation, networking, and physical processes”, ranging from minuscule (e.g. pacemakers) to large-scale (e.g. national power-grid) and typically involving feedback, \cite{raj10, damm16, toerngren16, Platforms4CPS2018}. While CPS have been around at least since the late 1970s (depending on how you interpret the term), they are today provided with entirely new capabilities due to improvements in various technologies, ranging from sensors, communication, computation, and artificial intelligence (AI) and machine learning (ML), algorithms, to new materials, batteries, and additive manufacturing.  
Corresponding trends for CPS include operation in more complex environments, higher levels of automation, electrification, and CPS-cloud and development/operation integration. 
This paves the way for unprecedented market opportunities, leading to CPS deployment in more open environments in all kinds of application domains such as transportation, manufacturing, healthcare, and smart cities. 
This can be seen as letting the “robots out of their cages”, as exemplified with automated driving and co-bots (robots collaborating with humans~\cite{BOZHINOSKI2019150, damm16, toerngren16}). 

\begin{figure}
	\centering
	\includegraphics[width=0.75\columnwidth]{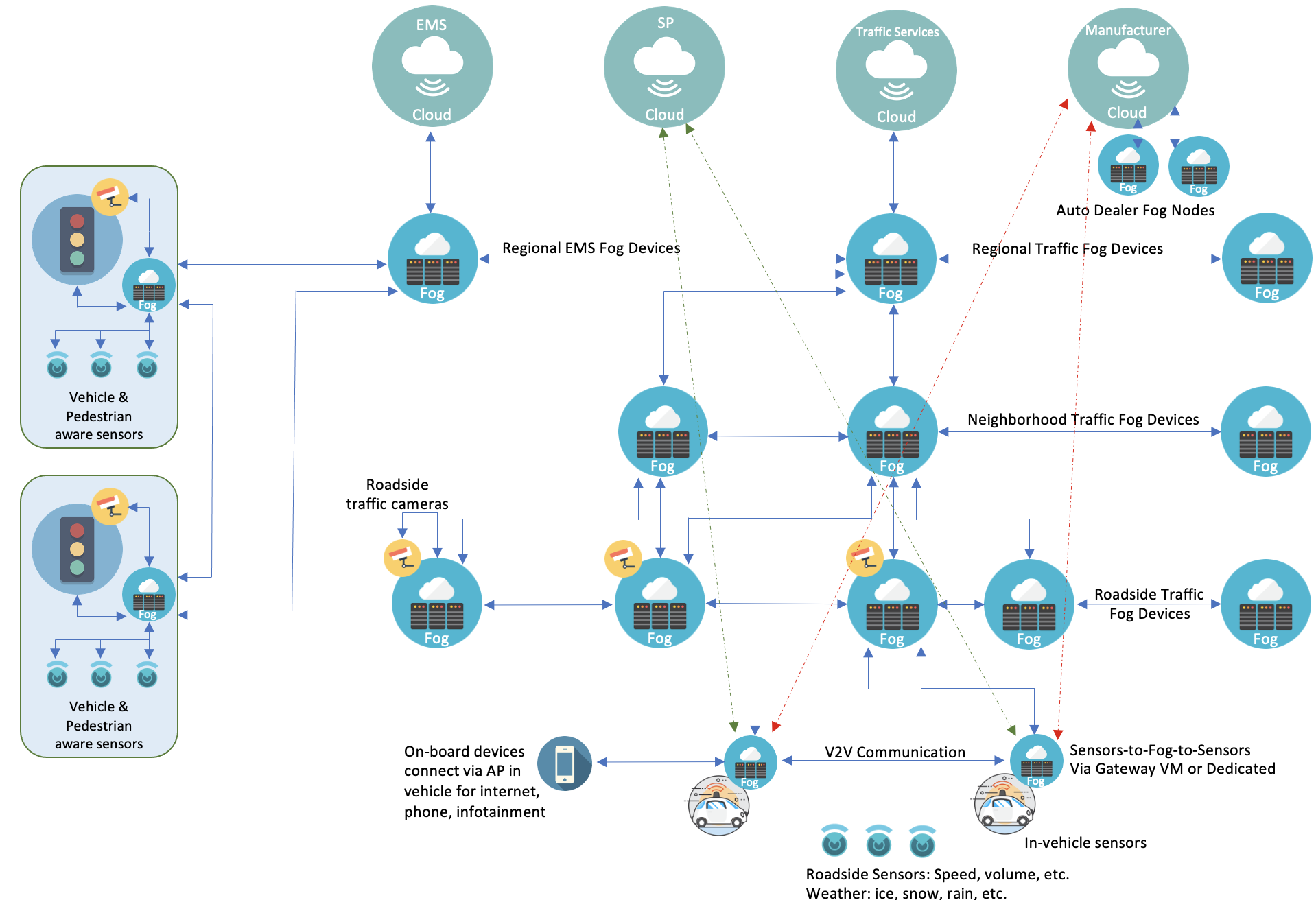}
	\caption{OpenFog architecture in ITS scenario (adapted from OpenFog Report \cite{OpenFog.2017})}
	\label{fig:ITS_tasks}
\end{figure}

CPS are often associated with critical applications where failures may jeopardize lives or where the lack of availability -- for example of infrastructure and manufacturing -- may have severe cost and/or safety implications. 
As an example of an integration between CPS and edge computing, Figure~\ref{fig:ITS_tasks} depicts a scenario with an intelligent transportation system (ITS) following the OpenFog reference architecture \cite{OpenFog.2017}. This ITS scenario brings the opportunity to examine the interactions among fog domains and cloud domains such as element management systems (EMS), service provider (SP), metro traffic services, and system manufacturer clouds.
By leveraging fog architecture, the strict requirements of this ITS applications can be accommodated. For instance, fog computing can be utilized to compute tasks obtained from a traffic control system or an autonomous vehicle. Therefore, the task can be performed in real-time to ensure the optimal and safe operations of the ITS.

\subsection*{Motivation}
The strong potential of future CPS comes along with new needs for computation, aligning with the strong drivers for edge computing. Future CPS are therefore likely to integrate edge computing in various forms, from device edge to the network edge (likely adapted to the needs and constraints of the respective domains), in essence adding a new tier to existing embedded systems and cloud computing on the cyber-side, This cyber-enhancement will enable to deploy new, enhanced and integrated cyber-physical capabilities.

Our interests lie in the applications of edge computing as part of Cyber-Physical Systems (CPS), where the introduction of edge-based CPS for critical systems requires an emphasis on trustworthiness and dependability. We note that both trustworthiness and dependability represent multifaceted properties, strongly related to CPS, security and human perception of trust, \cite{Dep}, \cite{nist16} \cite{Hanckock-Security}, \cite{NistGlossary}.
While the concept of dependability in its current form has been around for some 40 years, the concept of trustworthiness is now emerging as an umbrella term associated with how we, as humans, perceive the operation of increasingly advanced and complex CPS, see e.g. \cite{nist16, Platforms4CPS2018}, This use of the term is further underpinned by the relatively recent adoption of the term in the context of trustworthy AI, visible for example through the EU efforts towards trustworthy AI (see e.g. \cite{EU.2021.web} for an overview of work by the "High-level expert group on artificial intelligence" initiated by the European Commission). The technological shift is also implying that current methodologies and standards are partly inadequate to address challenges of future CPS (see e.g. \cite{Torngren2018}). This is clearly seen in the multitude of ongoing standardization efforts related to automated driving, cyber-security, and introduction of advanced perception and AI in the context of safety critical systems, see e.g. \cite{ISO-SAE-21434, SOTIF, UL4600, P2846, ISO-AWI-TS-5083} - reflecting different aspects of designing and assuring safety and security for highly automated CPS\footnote{For example, the so called SOTIF standard (ISO 21448) addresses safety aspects of machine learning and advanced perception systems for automated driving. As such it is representative for similar efforts also in other CPS domains (where also AI/ML and advanced perception systems are introduced, e.g. with similar ongoing work to extend \cite{IEC}. ISO 5083 addreeses safety for high levels of driving automation, including cybersecurity considerations} as well as in updated editions of traditional safety standards, see e.g. \cite{ISO26262}.

In this paper, we have chosen to emphasize the following attributes (or sub-properties) of trustworthiness: Safety, Security, and Predictability.
The rationale for this choice of attributes stems from industrial needs as derived in the TECoSA research center, \cite{TECoSA}; industry sees these three attributes as vital for introducing new edge-based CPS. Each of the attributes (safety, security and predictability) is facing new challenges as CPS expand to become edge-based, collaborative autonomous and "filled with" AI. Moreover, the mutual dependencies and trade-offs between these attributes also need to be explicitly considered. The selection of these attributes, and at this granularity also provides a delimitation of the scope of our survey.  We detail and elaborate on how we use Trustworthiness and its attributes as part of the survey in Section~\ref{sec:Classification}.

\subsection*{Contribution}
This paper, therefore, investigates the directions and concerns for the use of edge computing in CPS that need to be trustworthy. Considering the strong drivers and the relative novelty of the field, it becomes important to understand the specific requirements and characteristics of edge-based CPS, and to ensure that research is guided adequately to address specific gaps. 

We present the results of a systematic mapping study \cite{Kitchenham07}, a kind of systematic literature survey, investigating the use of edge computing for CPS with a special emphasis on trustworthiness. 

The main contributions of this study are:
\begin {itemize}
\item \textit{A detailed description of the current research efforts in edge-based CPS} - relating to CPS domains, types of applications and system aspects, and the type of edge computing considered (MEC, fog computing and cloudlets).
\item \textit{An analysis on how those research efforts address trustworthiness in terms of safety, security and predictability} - including combinations of these properties, and their relations to various edge-computing concepts and applications.
\item \textit{An analysis on the research gaps found during this study} - including recommendations for future work directions.
\end {itemize}



We first review related surveys of edge computing, CPS, and overlapping studies in 
Section~\ref{sec:RelatedWork}. We use a mapping study (this type of systematic literature survey is described in Section~\ref{sec:method}) and 
a classification, to structure and characterize research literature in the intersection between edge-computing and CPS, as described in Section~\ref{sec:Classification}. We present the results in Section~\ref{sec:Results}, where a link to the data can also be found. Next, we discuss the findings, identify research gaps and treat validity in Section~\ref{sec:Disc}. Finally, we elaborate on future work and recommendations for research in Section~\ref{sec:FutureWork} and provide concluding remarks in Section~\ref{sec:Conclusions}.

\section{Related work}
\label{sec:RelatedWork}

To the best of our knowledge, no previous paper has provided a broader systematic literature survey on the connection between edge-computing and CPS research.
However, several studies have addressed fog-computing for specific CPS domains such as smart cities and Industry 4.0, and many literature studies were carried out in related areas of edge-computing and CPS. In the following, we briefly describe surveys with some relation to our survey and the specific perspectives they provide.
As elaborated in the following, the surveys indicate needs to further address trustworthiness related properties of relevance for the use of edge computing in CPS \cite{Gonzalez2016, tocze2018, sat17, Ahmed2017, Khan2019}. 

\subsection*{Surveys on Edge computing in CPS}
In \cite{10.1145/3057266} a survey of fog computing for sustainable smart cities is provided, revealing that (i) cloud/fog collaboration ("cloud companion support"), (ii) data analytics, (iii) multi-protocol support at communication level, (iv) mobility, and (v) security and privacy represent commonly addressed research topics in fog computing applications. The paper draws a conclusion that both IoT and fog computing are comparatively immature fields, motivating a need for a focus on platforms for testing, experimentation, and evaluation. The importance to support multiple communication and application-level protocols, privacy and security (including authenticity, confidentiality, and integrity), and distributed intelligence is highlighted.

The topic of fog computing in the context of Industrial Internet of Things (IIoT) and Industry 4.0 has received a lot of attention in the research literature. For instance, \cite{Basir2019FogCE} reviews fog infrastructure and protocols in IIoT applications. Several communication and networking challenges are treated including 
(i) energy efficiency (e.g. balancing quality of service with energy consumption), (ii) network throughput and storage capacity (dependent on decisions of where to use and store data), (iii) resource allocation and spectrum use (as a challenge for network performance with impact on many quality of service parameters), 
(iv) latency, dealing with real-time connectivity requirements and the end-to-end chain of networking and processing (with several issues affecting latency such as resource allocation, network architecture, and node storage and energy capabilities), and (v) cache enabled edge devices (to reduce the load on backhaul links, and with schemes for efficiently accessing data).

In \cite{CAIZA2020e03706}, research papers on fog computing in the context of Industry 4.0 are surveyed. Industrial IoT protocols and applications are examined in terms of their architecture, latency, security, and energy consumption, and the authors highlight several challenges with industrial fog computing. In a more recent survey, \cite{Cao2021SurveyEdgeEdgeCloudComputingAssisted} considers edge computing-assisted CPS from a similar perspective of quality-of-service optimization. They define a series of critical challenges including latency, energy consumption, security and privacy, and system reliability.
In addition to classifying studies into these categories, they also summarize mechanisms for addressing them.

In \cite{Sitton.2019}, the authors conduct a survey on four edge computing reference architectures proposed by Intel-SAP \cite{intel}, FAR-Edge Project \cite{FAR}, Edge Computing Consortium \cite{ECC}, and the Industrial Consortium for Industry 4.0 \cite{IIC}. The aforementioned reference architectures are all based on a three-layer model for edge computing, which integrates all layers to process the service. We note that these reference architectures focus on edge computing for industrial environments, increasing the importance of ensuring e.g. system reliability and security. Although these four reference architectures have contributed partially to trustworthiness attributes with an emphasis on security, very few focused on other trustworthiness attributes such as safety and availability for edge-based CPS. 



\subsection*{CPS surveys}

There are several directions of CPS literature surveys, focusing e.g. on applications of CPS, see e.g. \cite{Chen-2017}, specific CPS domains (such as manufacturing/industry 4.0 or electrical grid), e.g. \cite{7740849, Lu-2017}, or on specific properties, such as security~\cite{10.1145/3313150.3313228}.



Literature surveys of CPS highlight connectivity, IoT, big data, and cloud interactions. Specific mentioning of fog or edge computing appears to be relatively rare and the surveyed literature generally focuses on technical or methodological aspects applicable to CPS in distributed computer system settings such as interoperability and performance. Challenges highlighted by \cite{Lu-2017, Chen-2017} include complexity of CPS, interoperability, cybersecurity, safety, dependability, and energy consumption.

CPS are often associated with critical applications; this is well recognized in CPS roadmaps and research challenge formulations~\cite{SRA-ECSEL-19, Platforms4CPS2018, BOZHINOSKI2019150}. Similarly, the increasing complexity of CPS with connectivity, collaboration, and more advanced algorithms including artificial intelligence and deep learning, poses both new opportunities and challenges~\cite{SRA-ECSEL-19, Platforms4CPS2018}. In the context of critical applications, this is especially true for properties such as security and safety, which face new challenges with new attack surfaces and faults/failure modes with complex behaviors and interactions in more open environments, requiring new approaches for system development, operation, and maintenance, see e.g. \cite{raj10, damm16, SRA-ECSEL-19, Platforms4CPS2018}. 

Another aspect is the increasing level of automation of CPS. NASA in  \cite{NASA-AssuringSafety} provides a comprehensive survey on safety assurance of increasingly autonomous systems. They identify open challenges regarding, (i) methodologies for safety assurance (e.g how do we go about designing and reasoning about the safety of autonomous CPS, and in providing automated reasoning to assist developers?), (ii) architecting autonomous systems to support assurance with an emphasis on pervasive monitoring, (iii) dealing with human-autonomous CPS interactions, and (iv) considering ethics for autonomous systems.  

These findings are also supported by comparisons of related agendas and roadmaps, see e.g. \cite{Platforms4CPS2018} and by the NIST CPS architecture framework \cite{NIST-CPS}.
The NIST framework was developed based on consultations with experts. It resulted in the identification of key life-cycle phases and aspects of CPS, with the aspects as representing 
groupings of cross-cutting concerns of relevance for one or more system stakeholders. Examples of key aspects identified include human-CPS interaction, trustworthiness, timing, data, and composability. The Platforms4CPS survey of agendas and roadmaps provided recommendations that address research, innovation, societal, legal, and business challenges related to CPS.
Particular emphasis was placed on trust-related concerns and CPS edge computing was highlighted as a specific research challenge.

\subsection*{Edge computing surveys}

Surveys of the various flavours of edge computing include those focusing on characteristics and requirements, e.g. \cite{sat17, Gonzalez2016, Khan2019}, resource management, 
\cite{Mao2017, tocze2018}, reference architectures \cite{Sitton.2019}, or specific technological instances such as multi-access or mobile edge computing, \cite{Mao2017, Taleb-7931566} and fog computing, \cite{7867731, Gonzalez2016, 8100873}.

In \cite{Khan2019}, a comprehensive survey of literature on edge computing paradigms is presented, providing characteristics of edge computing systems including fog computing, cloudlets, and mobile edge computing. Based on the survey, requirements for enabling edge computing systems are summarized, including availability, reliability, and security. Low-cost fault-tolerance and security are put forward as open challenges. Additional application and challenge perspectives are provided by \cite{sat17}, considering requirements for IoT applications including wearable cognitive assistance and favorable properties of edge-based realizations including availability, privacy, and latency. Challenges ahead, including complexity, security, and viable business models, are discussed. A previous survey on mobile edge computing of the same authors \cite{Ahmed2017} discusses requirements and challenges. Requirements mentioned for edge computing include reliability, scalability, resource management, security, interoperability, and business models. Open challenges put forward include seamless edge execution handover, eco-systems and business models enabling collaboration, lightweight security and privacy, and real-time data processing at scale.
\cite{Gonzalez2016}	investigates fog computing and specifically highlights challenges related to performance, security, and governance.

In \cite{8100873}, a survey on fog computing,  
focusing on algorithms and architectures, is presented. 
The paper describes expectations and the suitability of fog computing for future Tactile Internet applications involving physical tactile experiences and remote real-time control, with example applications in telesurgery,
and vehicle platooning. 
Requirements of connectivity and latency are elaborated for such applications, including expected end-to-end latencies of \SI{1}{\milli\second} or less and a maximum of \SI{1}{\second} outage per year. Challenges discussed include the design of higher layer APIs and protocols on top of lower layer protocols (e.g. provided by 5G), algorithms for tactile applications as well as novel resource management algorithms.

  
A complementary perspective is taken by \cite{tocze2018}, by focusing on resource management independent of the type of edge computing system.
The findings indicate a relatively low coverage of non-functional\footnote{Sometimes such properties are also referred to as "extra-functional", meaning that they specify e.g. how well or how to scale one or more functions.} properties in the literature; those covered in the paper include response time, energy, availability, and resource efficiency (in terms of resource utilization). Another study \cite{Mao2017} also focuses on resource management but in the specific context of mobile edge computing, emphasizing joint radio-and-computational resource management. Privacy and energy-related issues are included.

In \cite{bakhshi_dependable_2019}, a systematic literature review was conducted on dependability and fog computing. This study provides an overview of the current state of the research, analyzing dependability attributes, sources of threats, and threat management techniques. The authors identified reliability and availability as the most studied dependability attributes. Node failure and link or path failure were the main sources of failure reported in the literature. The study also focused on the means applied to ensure dependability, identifying redundancy techniques as the most common methods. The relation between safety and security in the solutions proposed for fog computing was also considered, finding very few studies that address both topics. Finally, it identified certain research gaps, e.g. reintegration after fault recovery in distributed systems. 

The survey \cite{renSurveyEndEdgeCloudOrchestrated2019} presents an overview of the emerging edge computing paradigms (fog, MEC, cloudlet), from the perspective of orchestrating the storage and computing resources of end-devices, edge servers and the cloud, which the authors call \textit{end-edge-cloud orchestration}, and the paradigms are compared and evaluated in terms of offloading, caching, security and privacy.
In the study, the authors also argue that transparent computing%
\footnote{An extension to the classical von Neumann architecture, where the lowest layers of a computer system is extended over a network. By leveraging block-streaming and just-in-time compilation, data and instructions can be fetched and executed over a network instead of the local bus.} 
shares this commonality, and thus include it in the survey. However, this study appears to be unique in this regard, and we chose to not include transparent computing in our final categorization.

Furthermore, in the ambitiously titled survey \cite{Yousefpour2019AllOneNeedsKnowFog}, "all one needs to know" about edge computing paradigms, includes taxonomies over fog, MEC and cloudlet architectures, and evaluations of their quality-of-service, security, RAS (reliability, availability, survivability) and management, to name a few objectives. The paper concludes by identifying challenges and research directions, several of which relate to trustworthiness aspects including resilient fog system design (considering reliability and availability, and trade-offs w.r.t. latency, throughput and security), fog system service level agreements, and various further security aspects such as trust and authentication in Heterogeneous Fog Systems.




\section{Method}\label{sec:method} 
A systematic mapping study is a well-established methodology from the Software Engineering research community that provides a structured classification of papers (a map of the field), where the classification relates to the corresponding research questions \cite{PETERSEN2008}.  
Systematic mapping studies are used by researchers based on established existing guidelines and well-defined steps. 

\begin{figure}[ht]
    \centering
    \includegraphics [scale=.8]{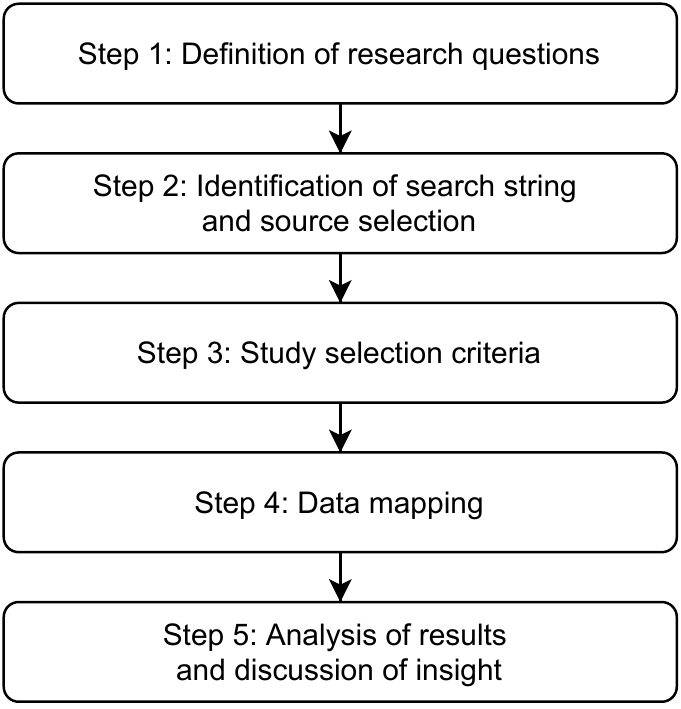}
    \caption{Workflow of the research method process}
    \label{fig:workflow} 
\end{figure}

Our systematic mapping study follows the guidelines presented in \cite{PETERSEN20151, Kitchenham07}. 
The process is adapted from \cite{abbaspour_asadollah_10_2017}, and
consists of the following main steps as presented in Figure \ref{fig:workflow}.
The details of each step are presented in the next subsections.

\subsection*{Definition of research questions (Step 1)} \label{sec:RQs}
The main goal of this study is to investigate the use of edge computing for Cyber-Physical Systems (CPS) and to find research gaps. 
This goal is refined into the following research questions (RQs):
\begin{description}
\item[RQ1:] How are edge computing solutions used for, or considered together with CPS in research? 
\textit{Objective}: to identify the areas where edge computing is being investigated, which technologies are used, and how trustworthiness is treated in the context.
\item[RQ1.1:] Which CPS domains are in the focus of edge computing?
\item[RQ1.2:] Which edge computing solutions are used for CPS?
\item[RQ1.3:] Which attributes (or aspects) of trustworthiness are addressed within edge computing for CPS?

\item[RQ2:] What types of applications within CPS are being treated with edge computing?
\textit{Objective}: to identify which application types are using edge techniques in the field of CPS, and to identify research gaps in it.

\item[RQ3:] What type of research is being conducted within edge computing for CPS? 
\textit{Objective}: to characterize what the individual studies emphasize, in terms of research contribution.

\item[RQ4:] What other factors  are influencing the development of edge computing for CPS?
\textit{Objective}: to analyze other trends in the development of edge computing technology within CPS.

\item[RQ4.1:] What classes of Artificial Intelligence (AI) are being used in edge-based CPS context?
\item[RQ4.2:] What type of edge computing solutions for CPS consider energy efficiency?

\end{description}

\subsection*{Identification of search string and source selection (Step 2)}
The main focus of this section is to identify a search string
and the selection of database sources to apply the search to achieve both a good coverage of existing research on
the topic and a manageable number of studies \cite{Kitchenham07}.

\subsubsection*{Search string}
A relevant search string should be able to return research works from the databases that address the study's RQs. For our research, we are interested in the intersection of two \textit{domains}, namely those of edge computing solutions, and cyber-physical systems.
To characterize each domain, synonyms of the main keywords and terms related to the respective domains are combined using the logical \textit{OR} operator. 
The following list includes the terms used to define each of the domains.

\begin{multicols}{2}
\paragraph{Domain A}
\begin{itemize}
    \item Edge computing
    \item Fog computing
    \item Cloudlet
\end{itemize}
\paragraph{Domain B}
\begin{itemize}
    \item Cyber-physical systems
    \item CPS
    \item Industry 4.0
\end{itemize}
\end{multicols}
The wildcard character "$*$" is used to provide results with and without hyphenation.
To compose a search string for such an intersection, the logical operator \textit{AND} is used, to return studies that belong to both sets.  The final search string is shown in \autoref{tab:string}. 

\begin{table}[ht]
\centering
\caption{The final search string}
\begin{tabularx}{0.8\columnwidth}{| >{\arraybackslash}X |}
 \hline
(``edge computing'' OR ``fog computing" OR cloudlet) AND \\
( ``cyber*physical" OR CPS OR ``industry 4.0” )
  \\
 \hline
\end{tabularx}
\label{tab:string}
\end{table}

\subsubsection*{Source selection}
In order to find the existing relevant occurrences for this topic, two scientific online digital libraries were chosen: \textit{IEEE Xplore Digital Library}\footnote{IEEE Xplore Digital Library [Online]. Available: \url{https://ieeexplore.ieee.org/Xplore/home.jsp}},
\textit{ACM Digital Library}\footnote{ACM Digital Library [Online]. Available: \url{https://dl.acm.org/}}.

The presented search string is used to query the studies from the sources with the necessary adaptations made in the syntax. The query resulted in a total amount of 667 
{candidate} studies. The total number of retrieved studies from each database is shown in \autoref{tab:tab3}.

\begin{table}[ht]
    \centering
    \caption{Number of studies retrieved from each library catalog}
    \begin{tabular}{l c}
        \toprule
        \textbf{Digital Library}    & \textbf{Search Results}  \\
        \midrule
        ACM Digital Library         & 338  \\
        IEEE Xplore Digital Library & 329 \\
        \midrule
        \textbf{Total}              & \textbf{667} \\
        \bottomrule
    \end{tabular}
    \label{tab:tab3}
\end{table}

\subsection*{Study selection criteria (Step 3)}
This step performs the shortlisting/selecting the relevant studies that are identified in the previous step based on some inclusion/exclusion criteria. For a study to be classified as relevant, it should meet all the inclusion criteria at once, and none of the exclusion ones. The inclusion criteria include all the studies referring to edge computing within the domain of CPS or Industry 4.0. The exclusion criteria determines which studies to be excluded: we excluded studies that are duplicates of other studies; studies that are not peer-reviewed, tutorial papers, and poster papers; survey studies are removed since they lie outside the scope of the mapping. Instead, the relevant survey studies are covered in Section~\ref{sec:RelatedWork}.

\begin{figure}[ht]
    \centering
    \includegraphics[scale=0.67]{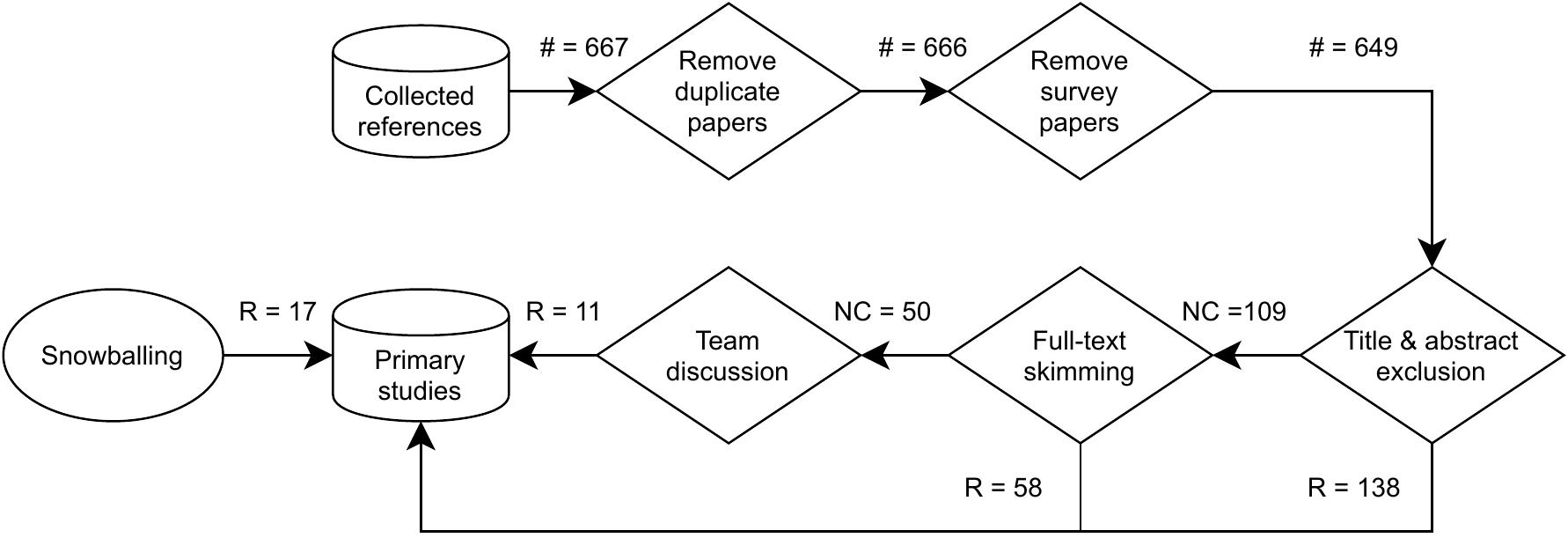}
    \caption{Overview of the study selection process}
    \label{fig:selectionprocess} 
\end{figure}

The selection process includes several steps and is detailed in Figure~\ref{fig:selectionprocess}.

After obtaining the total amount of studies from the automatic search, the first step to select the relevant studies begins with the removal of the duplicates. We used Zotero\footnote{https://www.zotero.org/} to identify these duplicates
and remove them. This is an open-source tool and is widely used. In the next step, all survey studies were removed, leaving the process with 649 studies after this step.
For the next step, "Title \& abstract exclusion", the studies were divided amongst the participating researchers/authors for a review of the paper titles and abstracts, resulting in a flagging of papers as "Relevant" (R), or "Not clear" (NC), or "Not relevant". 
A study was marked as relevant if it met all the inclusion criteria and none of the exclusion criteria, or as not-relevant if it lacked one of the inclusion criteria or met at least one of the exclusion criteria, or as not-clear if there were uncertainties arising from the title and abstracts review. 
All studies flagged as NC were examined closely by means of full-text skimming. Any subsequent studies that remain NC were brought to discussion.

\subsubsection*{Snowballing}

After identifying the primary studies, the snowballing aimed at identifying additional papers by using the references of the already identified relevant papers. This was accomplished automatically using the free and open-source reference management software Zotero with the AI-backed search engine Semantic Scholar (S2), by integrating their web APIs\footnote{Git repository [Online]. Available: \url{https://gits-15.sys.kth.se/nilsjor/zotero-s2-api}}.

The snowballing resulting in the retrieving of 4705 references. After removing duplicates, 3792 studies remained. These were then compared with the list from the original search
string, the 667 papers shown in Figure \ref{fig:selectionprocess}, reducing the number to 3709. The resulting list is finally filtered using the same search string described above; the studies had to include a term related to CPS and a term related to edge computing. This process reduced the final results to 24 references. Finally, the content of the studies was analyzed, adding a total of 17 new sources from the snowballing (see left bottom part in Figure~\ref{fig:selectionprocess}).

At the end of the Study selection process, there were a total of 224 studies flagged as Relevant, and the data mapping step could begin. 

\subsection*{Data mapping (Step 4)}
The next step in the systematic mapping study is to establish how the {relevant} studies are to be classified, in order to attempt to answer the underlying research questions of the study. This section provides the definitions used in the classification scheme.

Beyond the general publication data like title, name of the author(s), and publication year, seven additional facets are considered. The first, being the \textit{research types} presented in \cite{Wieringa}, which are adopted unaltered in our study, like in \cite{abbaspour_asadollah_10_2017}. 
The others are \textit{CPS domain}, \textit{edge implementation}, \textit{application class}, \textit{trustworthiness}, \textit{artificial intelligence}, and \textit{energy efficiency}. 
The initial categorization is based on the acquired knowledge from discussions with experts from their respective fields. 
Next, weekly meetings were held to refine the categories further. 

We defined a well-structured form based on the classification scheme and the research questions in order to extract the data from the selected relevant studies. Microsoft Excel spreadsheet was used to organize and store the extracted information from each relevant study for subsequent analysis.

\subsection*{Analysis of results and discussion of insight (Step 5)}
The last step of the systematic mapping study process is results analysis, where the map of the field is produced from the relevant studies and then a comprehensive analysis of the studies is performed to address the research
questions. We used multiple methods to produce the maps such as bubble charts, pie charts, and line graphs. Produced maps and the derived analysis
are presented in Section \ref{sec:Results}, and a discussion of the insights is presented in Section \ref{sec:Disc}.

\section{Classification Schemes Definition} \label{sec:Classification}



This section explains all the classifications in detail.



\subsection*{Research type, adopted from \cite{Wieringa}}
\begin{itemize}
\item \textbf{Validation research}	concentrates on investigating a proposed solution, which is novel and has not yet been implemented in practice. Investigations are carried out systematically, i.e., prototyping, simulation, experiments, mathematical systematic analysis, and mathematical proof of properties.
\item \textbf{Evaluation research}	focuses on evaluating a problem or an implemented solution in practice, i.e., case studies, field studies, and field experiments.
\item \textbf{Solution proposal}	provides a novel solution for a problem or a new significant extension to an existing technique.
\item \textbf{Philosophical paper}	describes a new way of looking at things by structuring in form of a conceptual framework or taxonomy.
\item \textbf{Opinion paper}	expresses the author's opinion whether a certain technique is good or bad.
\item \textbf{Experience paper}	sketches on the personal experience of the author, i.e., what and how something has been done in practice.
	
\item \textbf{Survey paper}	represents research where data and results are taken from other, already existing publications, where conclusions are drawn regarding trends, challenges, areas of interests, and future work, etc.
\end{itemize}

\subsection*{CPS Domain}
\begin{itemize}
\item \textbf{Telecommunication (Telecom)} includes communication infrastructures, wireless communication, applications of 5G mobile networks, etc.
\item \textbf{Healthcare} includes the technologies to monitor and give medical care to the patients in hospitals as well as customers outside the hospital. 
\item \textbf{Manufacturing} includes sensing, actuation, big data analysis, communication, control, and optimization of manufacturing systems. Typical CPS-related concepts include cloud manufacturing, industry 4.0, and digital twins.
\item \textbf{Infrastructure} includes smart buildings/homes and smart cities. The CPS technologies enable remote monitoring and control and hence have the potential to improve the safety, security, and energy efficiency in for example smart buildings. 
\item \textbf{Energy} encompasses energy-related considerations in for example smart grids, power plants, household electricity generation with renewable energy. Energy has been as one keep aspect related to sustainability. 
\item \textbf{Transportation} refers to different modes for transporting people and goods (cars, trucks, buses, trains, etc.). Major applications include for example autonomous vehicles, vehicle to X communications, and intelligent transportation systems.
\item \textbf{Other} referring to any other CPS domain mentioned.
\item \textbf{Not specified} -- relevant if the work is independent of any explicit CPS domain mentioning.
\end{itemize}

\subsection*{Edge implementation/concepts}
Among the many interpretations of edge computing, we find that “the edge” is given a different meaning as already mentioned in the introduction. We focus on the following implementations, or concepts, referring to edge computing which we understand as mainstream, see e.g.\cite{8016213, ELAZHARY2019105}.
 
\begin{itemize}
\item \textbf{Fog computing} can be seen as an extension of cloud computing introduced by Cisco Systems in 2012 \cite{Bonomi.2012}. 
It enables computing, storage, networking, and data management from the core of the network to its edges. Therefore, network performance can be enhanced given that the processes are not only executed in centralized cloud servers but also along the path to them. 
\item \textbf{Multi-Access Edge Computing} (MEC) is a platform that provides IT and cloud-computing capabilities within radio access network (RAN) in 4G and 5G, in close proximity to mobile subscribers \cite{Abbas.2018, Taleb-7931566}. Particularly, it is located on the network edge and provides computation capabilities and storage resources to nearby low energy, low resource mobile devices. 
\item \textbf{Cloudlet} is another direction in distributed mobile computing that shares many traits with MEC. Specifically, a cloudlet refers to a cluster of trusted computers with a strong connection to the Internet that is utilized by nearby mobile devices. Moreover, cloudlets are located in the middle tier of a 3-tier continuum, i.e., mobile device-cloudlet-cloud, and typically one hop away from mobile devices. The idea is to offload computation from mobile devices to a virtual machine (VM) based cloudlets located on the network edge. Therefore, cloudlets need infrastructure with VM capability \cite{Satyanarayanan.2009}. 
\item \textbf{Other} definitions of edge computing are used in the surveyed literature. Some research articles that we reviewed proposed their solution in terms of other related edge computing concepts such as mist computing, vehicular edge computing, etc). Thus, we classify those articles as "other".
\item \textbf{Not specified} has been assigned to studies without any explicit reference to any type of edge computing implementation.
\end{itemize}

\subsection*{Application class}
With application class we refer to the application or system aspect in focus for the research.
\begin{itemize}
\item \textbf{Resource management}	considers edge-based methods for handling system resources, such as scheduling, orchestration, migration, and distribution of computation, storage, etc.
\item \textbf{Collaborative CPS}  deal with systems where information is exchanged between several CPS, typically with edge computing infrastructure foreseen to support computation, for the purposes of collaboration.  
\item \textbf{Real-time application analytics}	concerns applications where edge computing can be leveraged to bring demanding real-time computation closer to the edge devices.
\item \textbf{Human-machine interaction}	are applications where edge computing can be used to provide low-latency feedback to human operators, such as augmented reality and cognitive assistance.
\item \textbf{Networked control systems}	include CPS with a closed-loop feedback control over the edge, and/or dynamical systems analyzed using control theory.
\item \textbf{Autonomous systems}	describes edge-device systems with a high degree of autonomy, even in the absence of other devices.
\item \textbf{System-internal monitoring}	denotes methods for measuring or otherwise detecting system characteristics, such as energy consumption, latency, or faults/failures.
\item \textbf{Software architecture}	is used to refer to structural/behavior arrangements and configurations of software and hardware components, e.g. related to concepts such as software-defined networking and blockchains.
\end{itemize}

\subsection*{Trustworthiness}

As mentioned in the introduction, we use the term trustworthiness as an umbrella property, focusing on the attributes of safety, security and predictability. 
This choice of attributes implies that the way we use the term comes relatively close to the concept of dependability, as "the ability to deliver service that can justifiably be trusted". 
Dependability encompasses the attributes of availability, reliability, safety, integrity, maintainability, and more recently, security, and in addition considers means to deal with these attributes (such as fault removal and tolerance) and "threats" to dependability (faults, errors, and failures), \cite{Dep}.

Trustworthiness is increasingly adopted in the context of CPS, see e.g. \cite{nist16, Platforms4CPS2018, EU.2021.web}. Trustworthiness as a concept, reflects an emphasis on the end properties of a system, where the resulting trust will stem from the integration of cyber- and physical parts, and their interactions with humans and other systems. 
This concept thus extends well beyond pure computing systems, and is suitable for CPS. 
Considering this adoption and usages of the term, trustworthiness has been our choice.
Given our emphasis on three trustworthiness, attributes, the corresponding classification is as follows. 

\begin{itemize}
\item \textbf{Safety} commonly concerns either an absolute or a risk-related property; we exemplify here with the latter interpretation, viewing safety as the "absence of unacceptable risk" from conditions that can lead to harm to people, property, or the environment, see e.g. \cite{IEC}. Safety considerations typically result in requirements on how a system is used and interacts with its environment, and on availability and reliability related properties of subsystems/components and their interactions. According to Firesmith, 
safety can be seen as the degree to which accidental harm is prevented, detected, and reacted to, \cite{Firesmith}. 
However, newer safety standards are beginning to highlight that harm may arise also from malicious intent and usage of a system, thus safety will increasingly rely on protection from attacks (security). For example, the ISO26262 edition from 2018, has the following statement: 
\textit{”5.4.2.3 The organization shall institute and maintain effective communication channels between functional safety, cybersecurity, and other disciplines that are related to the achievement of functional safety. EXAMPLE 1 Communication channels between functional safety and cybersecurity in order to exchange relevant information (e.g. in the case it is identified that a cybersecurity issue might violate a safety goal or a safety requirement, or in the case a cybersecurity requirement might compete with a safety requirement).”} \cite{ISO26262}. While it is important that such a point and reference to cyber-security is made, it is also evident that methodological guidance on how to accomplish this is urgently required, indeed representing a research topic that is drawing (and requiring much more)  attention as for example seen from the publications in recent Safecomp conferences.


\item \textbf{Security}, as opposed to safety, can be seen as the degree to which malicious harm is prevented, detected, and reacted to, \cite{Firesmith}. 
Security is in itself multi-attribute, taken for example to encompass authentication, authorization, integrity, confidentiality, and availability (see Chapter 4 in \cite{SecHandbook}).  
The increasing connectivity, and the introduction of edge-based CPS, provides both promises to deal with security attacks (by e.g. local monitoring and responses), but also exposes more attack surfaces, where attackers may leverage both the cyber-, physical, and humans dimensions (and their combinations) for attacks.

\item \textbf{Predictability} is a term traditionally associated with real-time computing  systems, referring to the ability to satisfy the timing requirements of critical tasks with some level of guarantee (depending on the static or dynamic nature of the systems), \cite{Stankovic}. Edge-based CPS will be dynamic in nature, with varying loads, partial failures or losses (e.g. loss of message packets), potential migration of computations, etc. To deal with real-time critical applications, a number of timing requirements may be relevant such as precise timing, age of data, and the corresponding detection of timing overruns, \cite{TorngrenJRTS}. This relates closely to the availability and resource management of end-to-end computation chains in an edge-based CPS. With predictability, we refer to both hard and soft real-time capabilities, including approaches that in some way address availability and resource management.  
\item \textbf{Combinations}	of several trustworthiness properties and their trade-offs will normally have to considered in edge-based CPS. We therefore also specifically searched for papers that considered combinations of these properties.
\item \textbf{None}	has been assigned to studies without any specific reference to a trustworthiness property.
\end{itemize}

\subsection*{Artificial intelligence}
In this paper, the primary interest is to understand which classes of artificial intelligence methods have been used in the context of trustworthy edge computing. Note that AI methods can either be applied to enhance the capabilities of edge computing infrastructure or used within applications on top of edge computing. For the sake of the present study, we have divided the AI technologies into the following classes:
\begin{itemize}
\item \textbf{Machine reasoning} refers to symbolic ontology-based methods working with declarative knowledge including logical reasoning.
\item \textbf{Machine learning}	includes numeric and symbolic learning methods including supervised, unsupervised, reinforcement learning, and combinations of those.
\item \textbf{Model-based methods} includes methods used for procedural knowledge processing, including state space exploration and AI-planning.
\item \textbf{Other} refers to methods that are not included in the categories above, such as evolutionary methods and game theory.
\item \textbf{None}	has been assigned to the studies without any specific reference to artificial intelligence.
\end{itemize}

\subsection*{Energy efficiency} This is a binary category where we have identified if a paper considers energy efficiency at the application level and/or the computing infrastructure. 

\section{Results} \label{sec:Results}
In this section we present the findings from the survey with a subsection for each of the four research questions as introduced in Section~\ref{sec:method}. The outcomes of the research questions are illustrated in the form of charts and/or graphs. 
The complete list of the relevant studies and their classifications in the study can be found online\footnote{Available in CSV-format: \url{https://zenodo.org/record/5112378}}.

\subsection*{RQ1: How are edge computing solutions used for, or considered together with CPS in research?}
\begin{figure}[ht]
    \centering
    \includegraphics[scale=1.0]{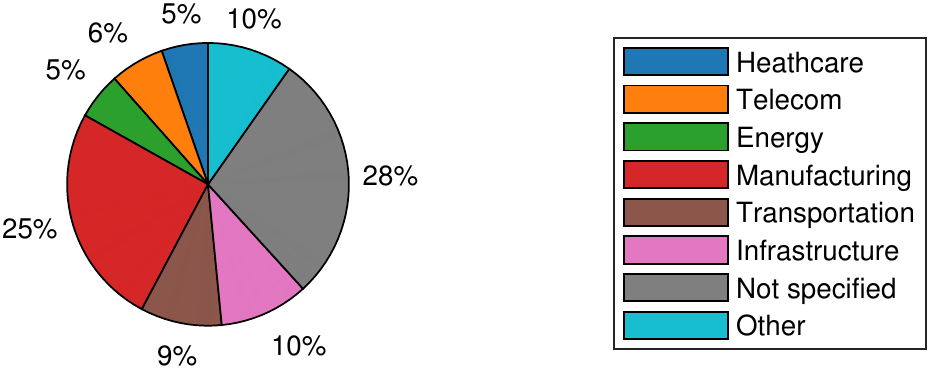}
    \caption{Distribution of the CPS domains considered in all relevant studies}
    \label{fig:PieChartCPS} 
\end{figure}

Figure~\ref{fig:PieChartCPS} shows the distribution of the CPS domains among the literature studied. The results show that the biggest group of studies address CPS in general, without specifying the domain. Among the ones that are related to a specific domain, manufacturing has the largest representation.

\begin{figure}[ht]
    \centering  
    \includegraphics[scale=1.0]{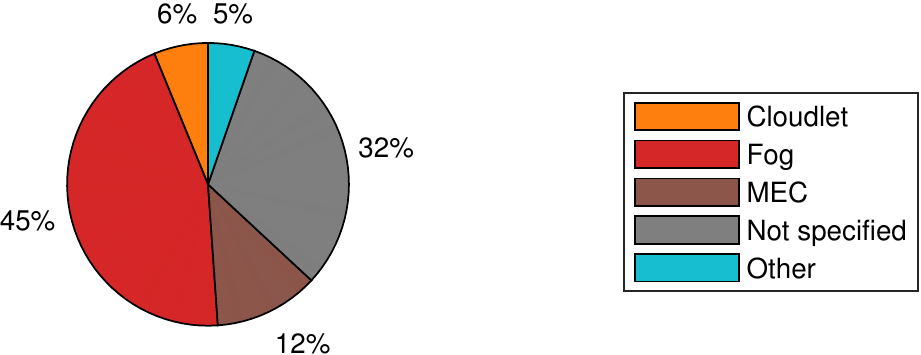}
    \caption{Distribution of edge implementations considered in all relevant studies}
    \label{fig:PieChartEdge} 
\end{figure}

The distribution of the different edge implementations is shown in Figure~\ref{fig:PieChartEdge}. Fog computing is overwhelmingly the largest category, covering nearly half of the total number of publications.

\begin{figure}[ht]
    \centering
    \includegraphics[scale=0.67]{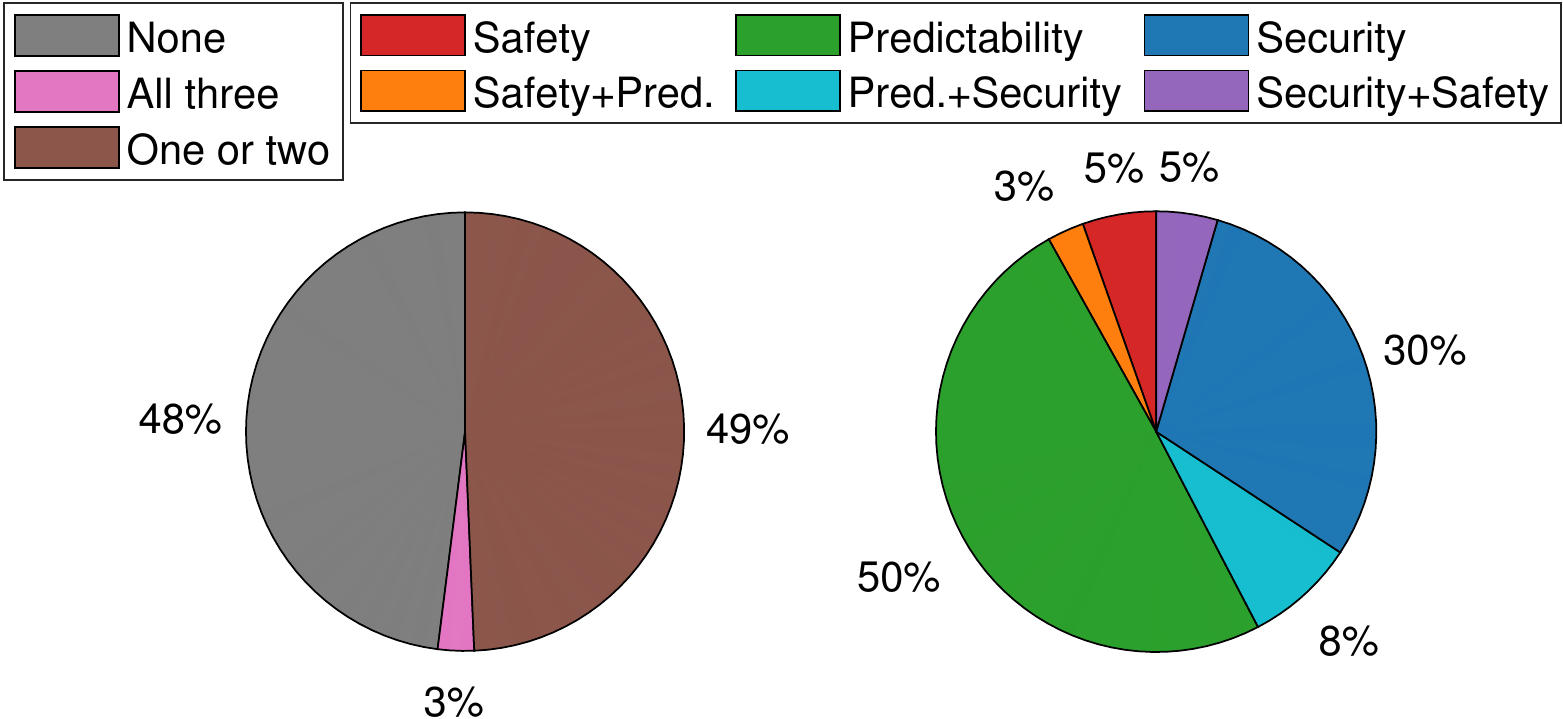}
    \caption{Trustworthiness concepts. The left chart shows how many distinct aspects of trustworthiness are considered, and the right chart shows a detailed breakdown of the "One or more" (brown) category.}
    \label{fig:PieChartTrust} 
\end{figure}

Finally, the distribution of the trustworthiness attributes is shown in Figure~\ref{fig:PieChartTrust}. The pie chart on the left shows that \SI{48}{\percent} of the studied paper do not consider any of the trustworthiness attributes, and only \SI{3}{\percent} of the studies consider all the trustworthiness attributes. The remaining \SI{49}{\percent} consider one or two attributes, and the chart on the right shows the break-down of publications in this category. Safety, security, and predictability are represented by a primary color and their intersections represent publications that mention two of them.

\begin{figure}[ht]
    \centering
    \includegraphics[scale=0.67]{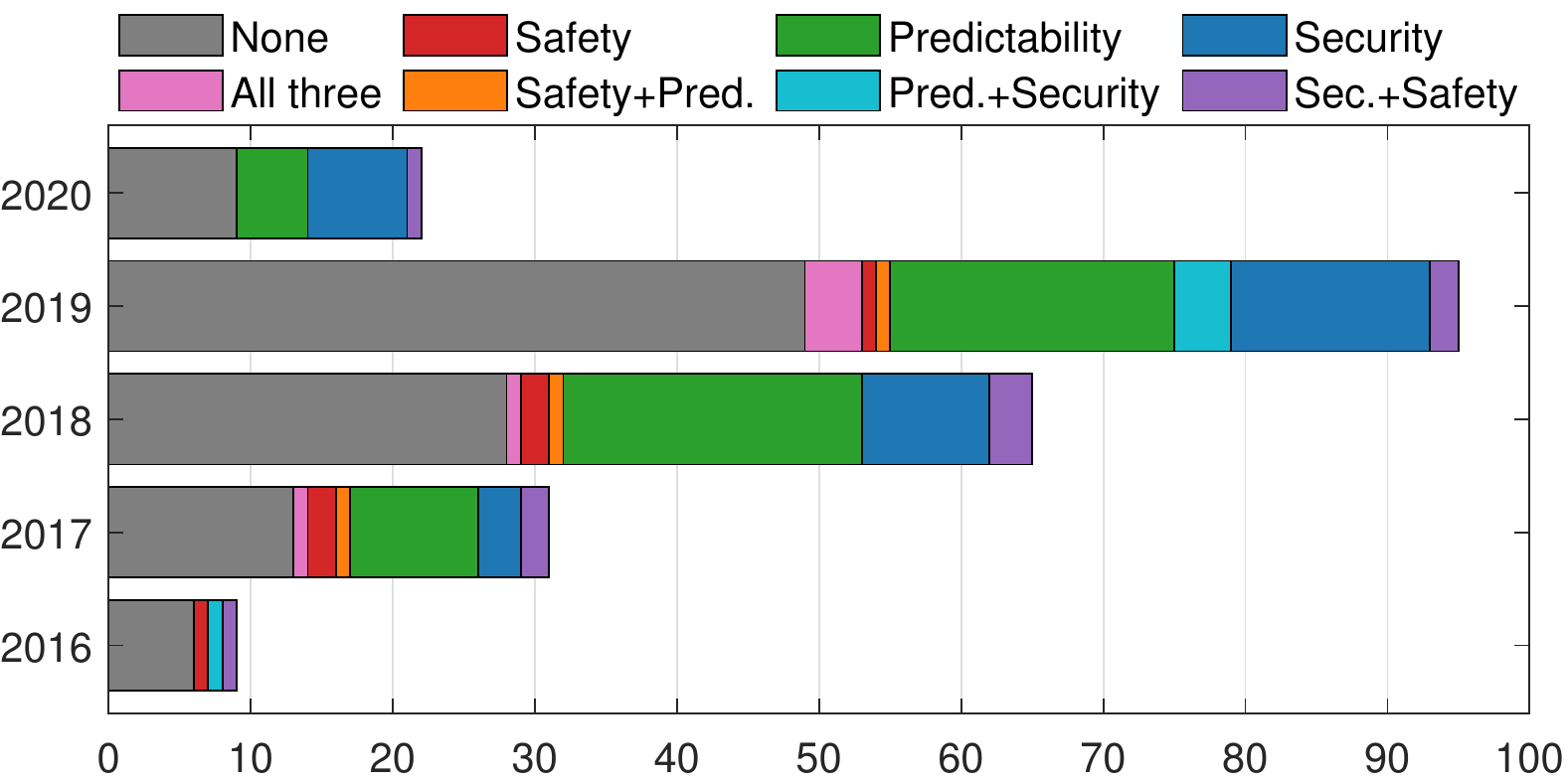}
    \caption{Trustworthiness categories over time. Note that data from the year 2020 is incomplete, owing to the timing of the study.}
    \label{fig:YearTrust} 
\end{figure}

An interesting aspect to consider is the evolution of the trustworthiness attributes in edge-based CPS over time, as seen in Figure~\ref{fig:YearTrust}. It is possible to observe how, despite the increase in the number of publications, the ratio of those that consider trustworthiness attributes remains relatively constant. The low number of publications in 2020 comes from the fact that the study began at the beginning of that year.

\begin{figure}[ht]
    \centering
    \includegraphics[scale=0.67]{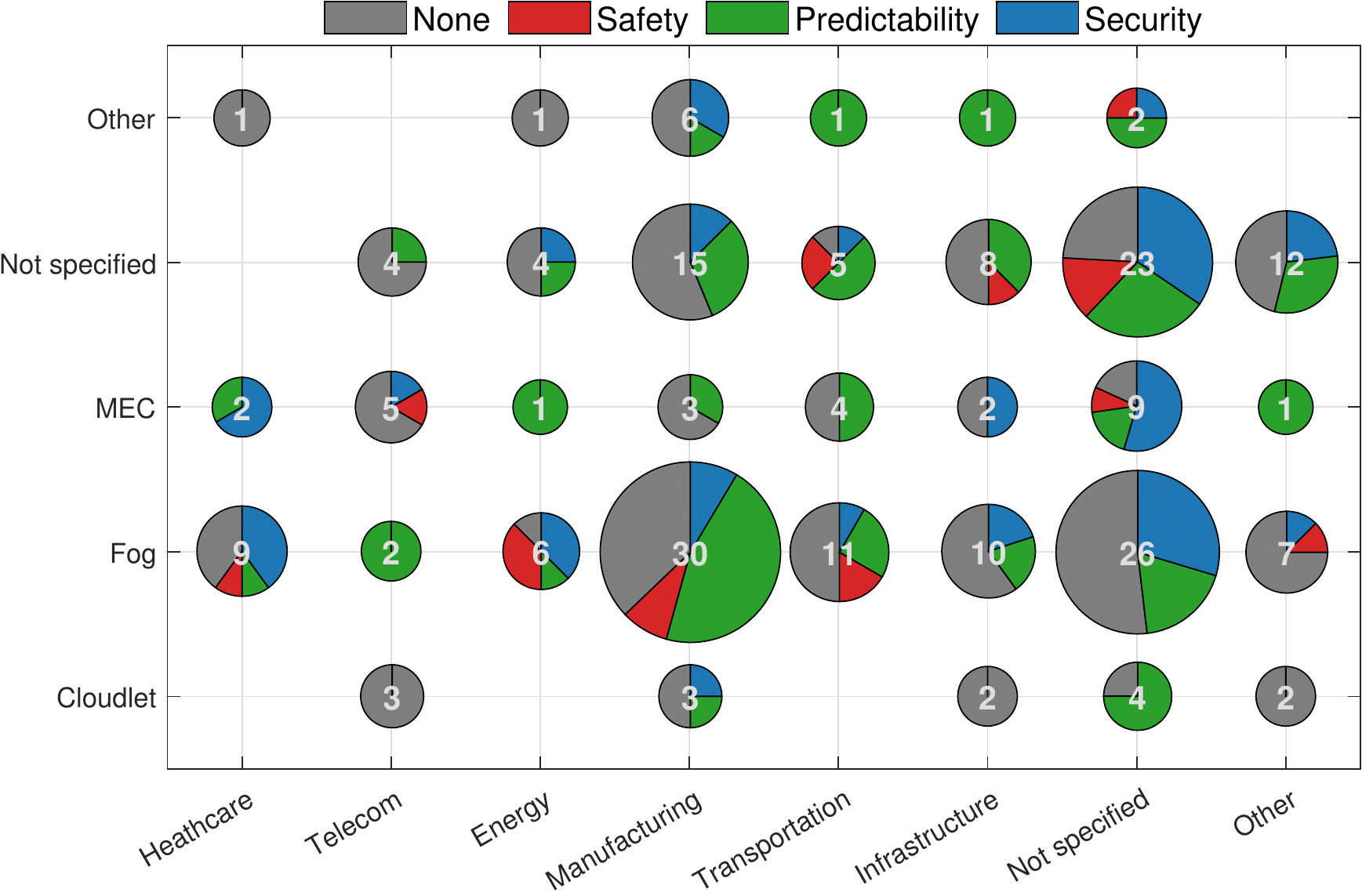}
    \caption{Relation between CPS domain, edge implementation and trustworthiness}
    \label{fig:CPSEdgeTrust} 
\end{figure}

Finally, the relations between CPS domain, edge implementation, and trustworthiness are presented in Figure~\ref{fig:CPSEdgeTrust}. The $x$-axis shows the CPS domains, the $y$-axis shows the edge computing implementations. In each intersection, the total number of publications is shown, as well as the relative coverage of the three trustworthiness attributes. In order to reduce the complexity of the pie charts, the intersections between the trustworthiness attributes are not shown. Instead, studies that address more than one attribute are counted once for each contribution. 

Figure~\ref{fig:CPSEdgeTrust} shows that manufacturing using fog computing as the edge implementation has received the largest attention. Some gaps are also noticeable, such as healthcare or energy applications using cloudlet-based edge computing.

\subsection*{RQ2: What types of applications within CPS are being treated with edge computing?}

\begin{figure}[ht]
    \centering
    \includegraphics[scale=1.0]{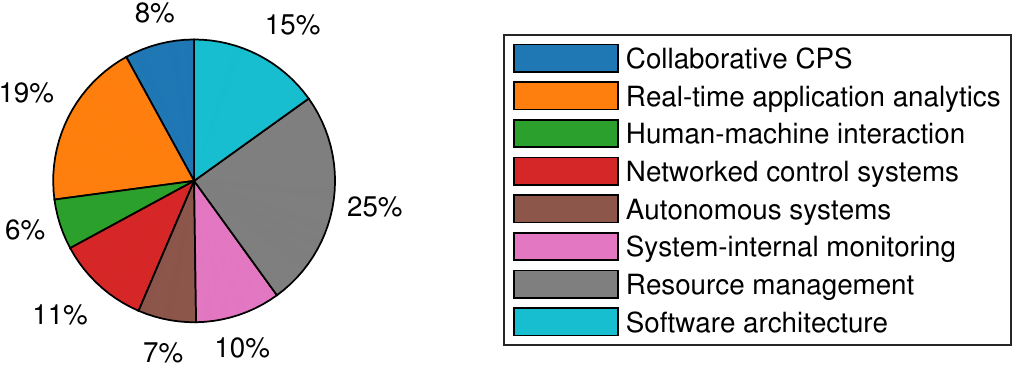}
    \caption{Distribution of application types considered in all relevant studies}
    \label{fig:PieChartApp} 
\end{figure}

Figure~\ref{fig:PieChartApp} represents the distribution of the application types, revealing that resource management and real-time application analytics are the most studied application types for edge-based CPS.

\begin{figure}[ht]
    \centering
    \includegraphics[scale=0.67]{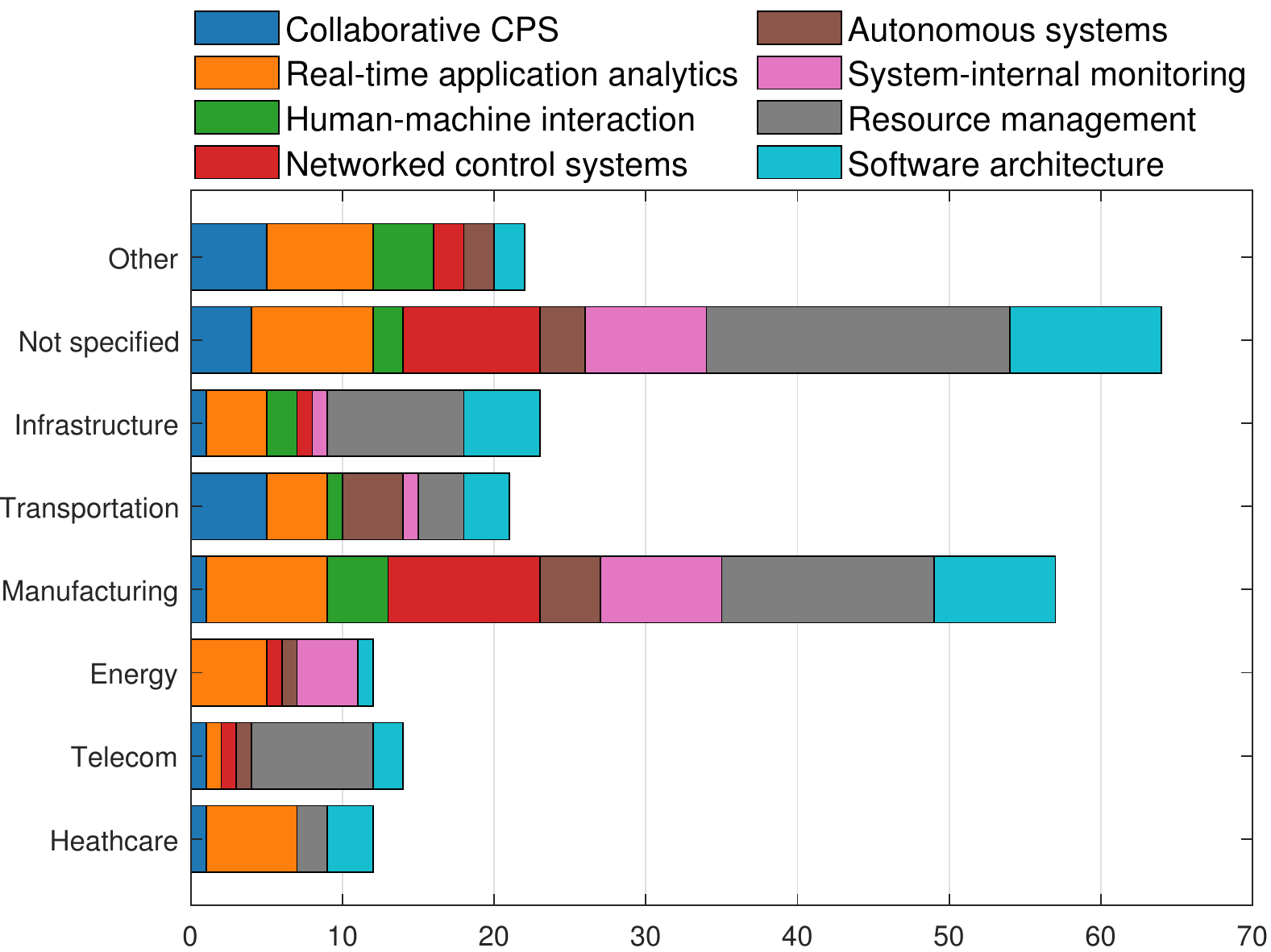}
    \caption{Applications distributed amongst the domains in all relevant studies}
    \label{fig:CPSApp} 
\end{figure}

Figure~\ref{fig:CPSApp} shows how the application types are distributed among the CPS domains. The $x$-axis represents the number of publications and the $y$-axis shows the different domains.

\begin{figure}[ht]
    \centering
    \includegraphics[scale=0.67]{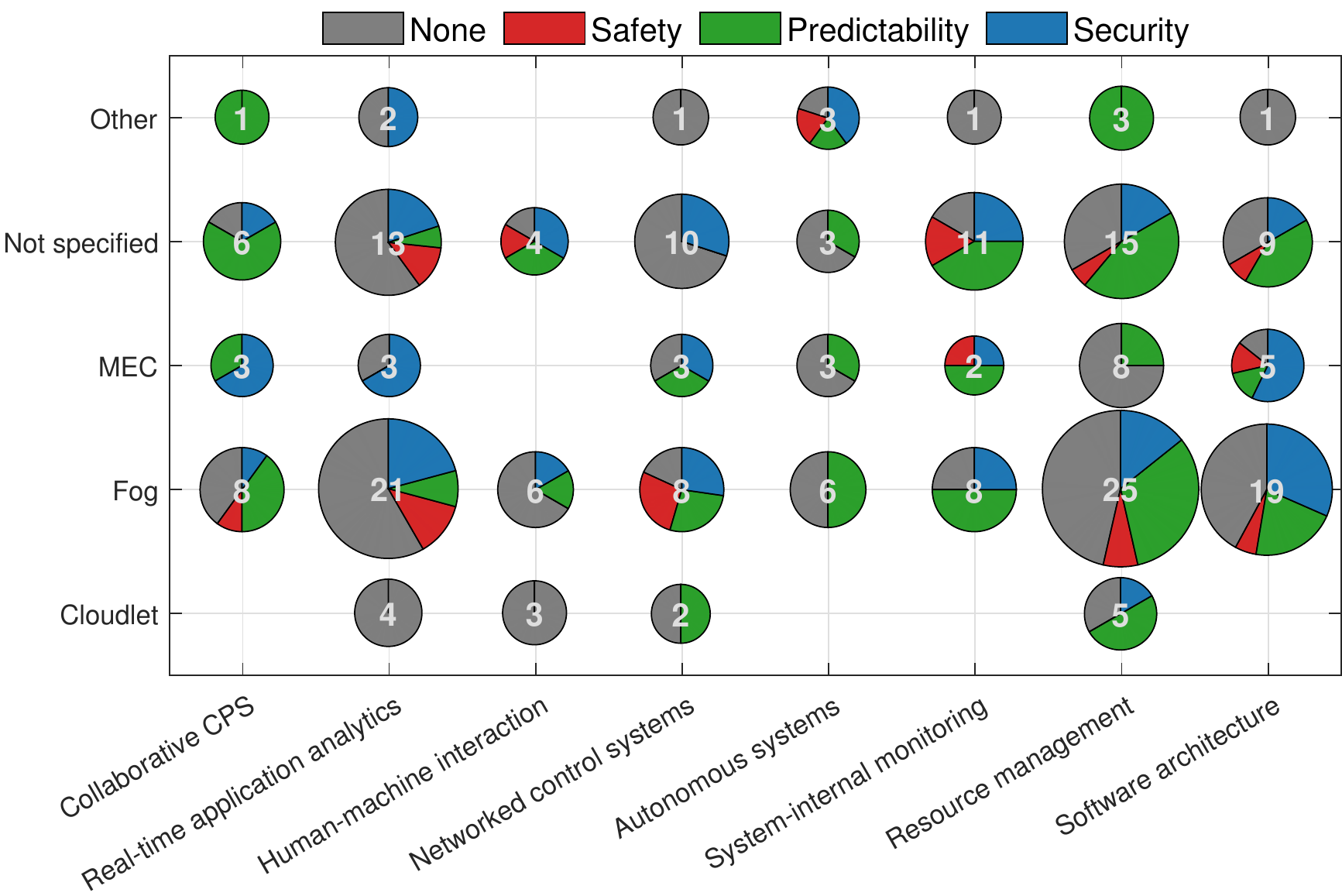}
    \caption{Relation between application type, edge implementation and trustworthiness}
    \label{fig:AppEdgeTrust} 
\end{figure}

Finally, the relation between application type, edge implementation, and trustworthiness is presented in Figure~\ref{fig:AppEdgeTrust}.
It should be mentioned that it closely resembles Figure~\ref{fig:CPSEdgeTrust}, but with the application class on the $x$-axis, rather than the CPS domain.
Real-time application analytics and Resource management using fog computing represent the largest groups. There are some research gaps, where none or very few publications have been found, e.g. human-machine interaction using MEC and system-internal monitoring using cloudlets. It can also be noticed that human-machine interaction received the least attention in the surveyed research.

\subsection*{RQ3: What type of research is being conducted within edge computing for CPS?}

\begin{figure}[ht]
    \centering
    \includegraphics[scale=1.0]{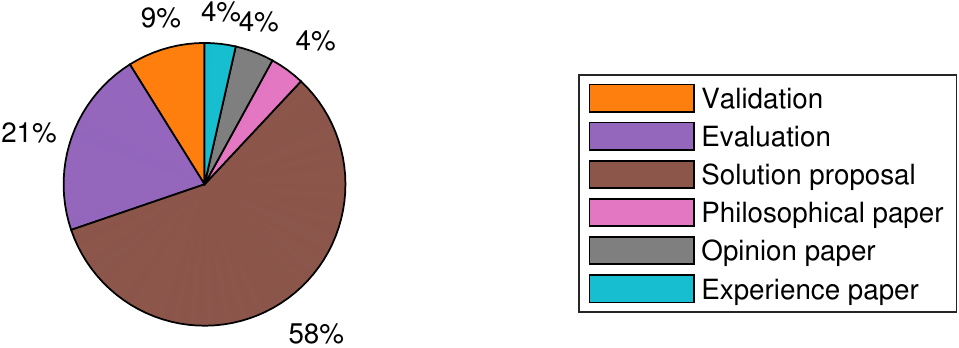}
    \caption{Distribution of research types in all relevant studies}
    \label{fig:PieChartType} 
\end{figure}

The distribution of the research types are shown in Figure~\ref{fig:PieChartType}. More than half of the studies have been classified as solution proposals, while evaluation and validation only represent \SI{21}{\percent} and \SI{9}{\percent} respectively. The remaining studies are shared among opinion, philosophical, and experience papers.

\begin{figure}[ht]
    \centering
    \includegraphics[scale=0.67]{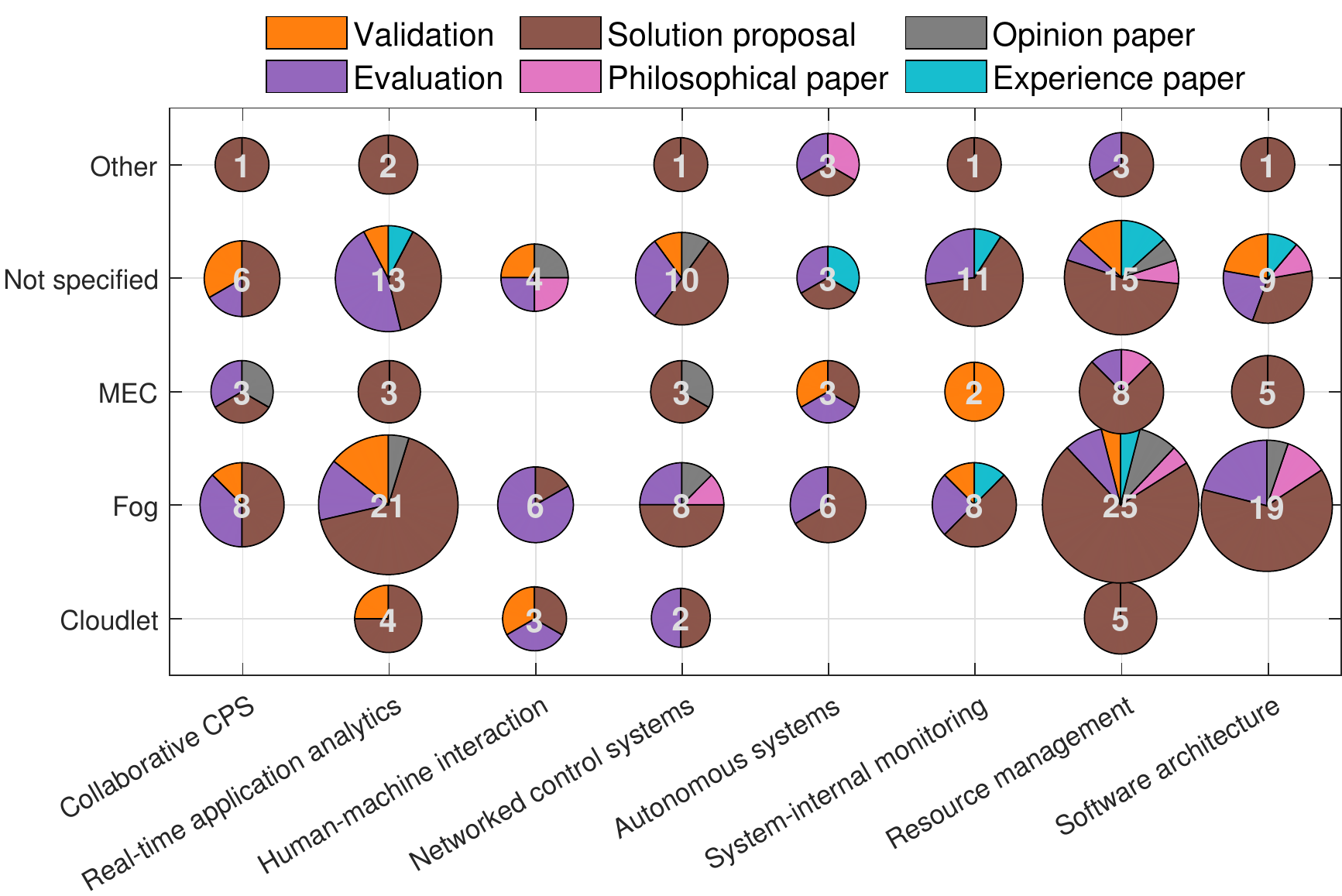}
    \caption{Relation between application type, edge implementation and research type}
    \label{fig:AppEdgeType} 
\end{figure}

Figure~\ref{fig:AppEdgeType} shows the relation between application type, edge implementation, and research type. Regarding the research type, it is visible that solution proposal is the predominant category in almost every group. Evaluation and validation tend to occupy the second and third positions, but there are quite a few groups where these categories are not present.

\subsection*{RQ4: What other factors are influencing the development of edge computing for CPS?}

\begin{figure}[ht]
    \centering
    \includegraphics[scale=1.0]{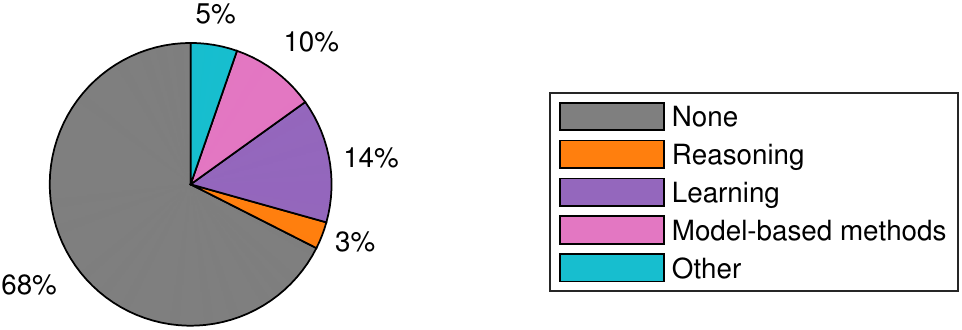}
    \caption{Distribution of AI methods considered in all relevant studies}
    \label{fig:PieChartAI} 
\end{figure}

The distribution of the AI methods used in the studies is shown in Figure~\ref{fig:PieChartAI}, illustrating that two-thirds of the publications do not mention any kind of AI. Among the studies that use AI, learning methods are the most common ones.

\begin{figure}[ht]
    \centering
    \includegraphics[scale=0.67]{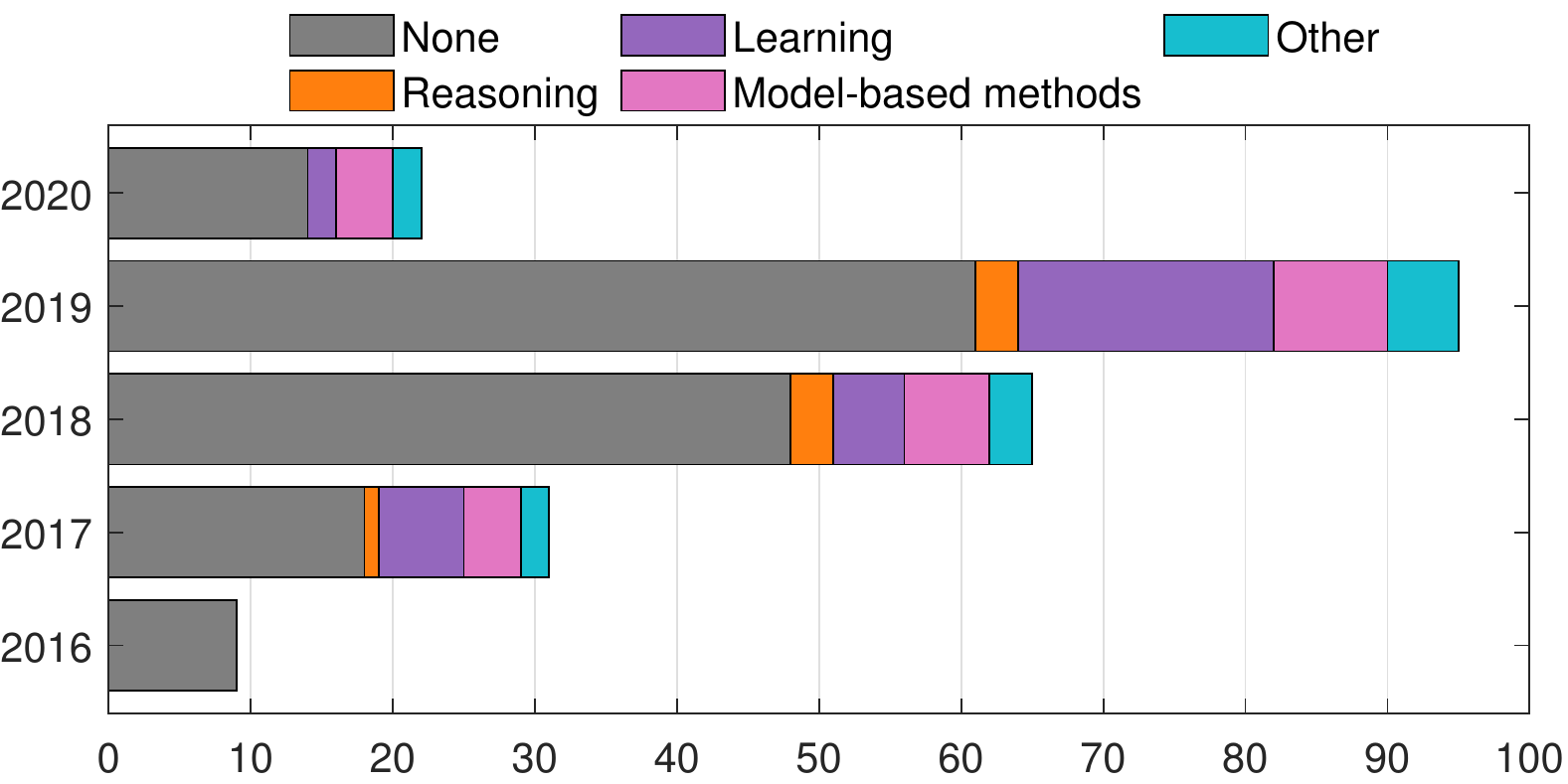}
    \caption{Distribution of AI over time. Note that data from the year 2020 is incomplete, owing to the timing of the study.}
    \label{fig:YearAI} 
\end{figure}

However, the evolution of those categories over time gives a slightly different picture, as shown in Figure~\ref{fig:YearAI}. It can be seen that the interest in learning methods within edge computing for CPS increased substantially for the year 2019.

\begin{figure}[ht]
    \centering
    \includegraphics[scale=0.67]{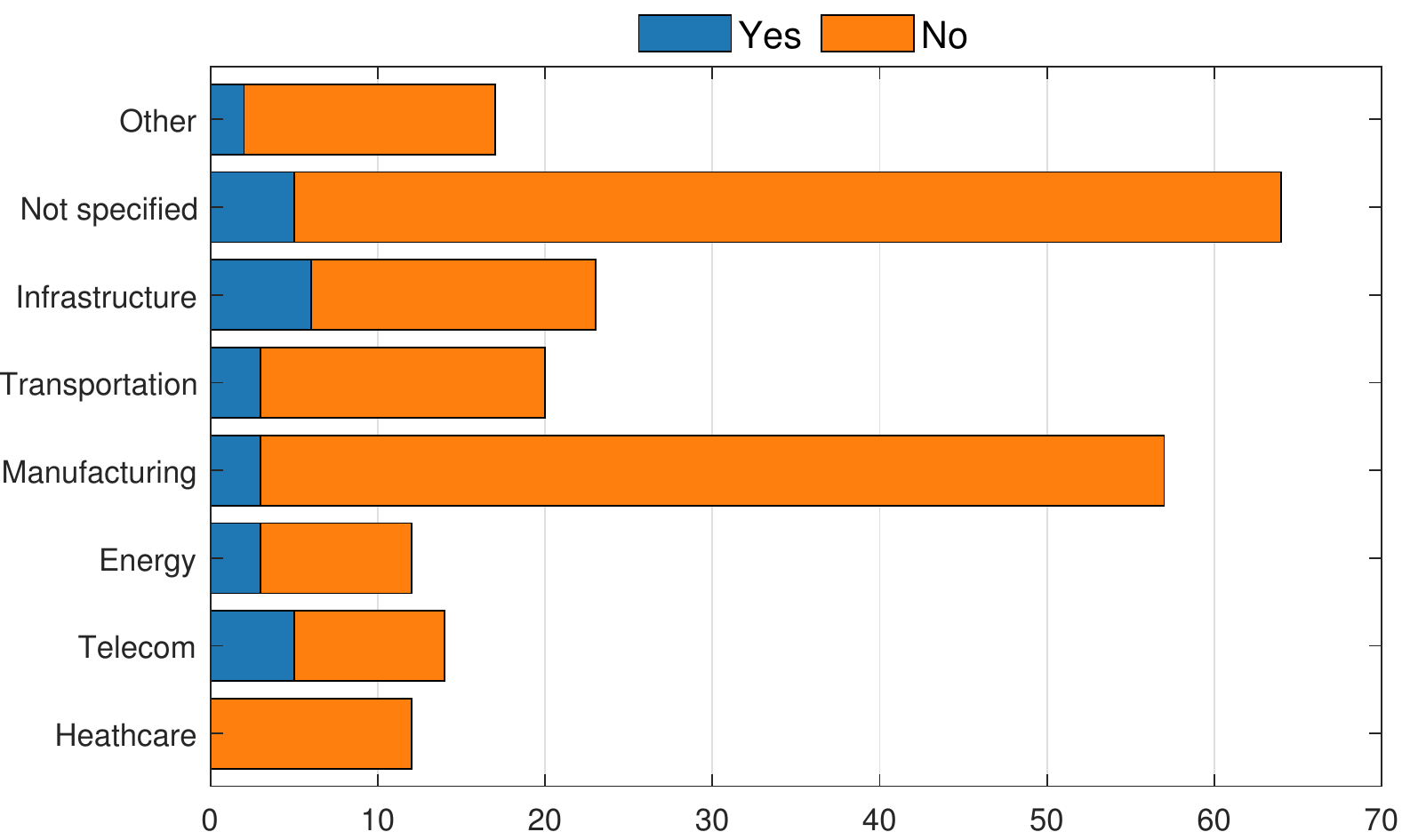}
    \caption{Energy efficiency as considered within the different domains in the relevant studies}
    \label{fig:CPSEnergy} 
\end{figure}

Finally, with energy efficiency chosen to represent sustainability, only \SI{9}{\percent} of the studies consider some aspect of energy efficiency. 
Figure~\ref{fig:CPSEnergy} shows the distribution of studies considering energy efficiency per CPS domain.


\section{Discussion}
\label{sec:Disc}


Our systematic mapping study, as well as the related surveys, clearly paints a picture of edge-based CPS as an emerging field that addresses multiple types of applications.
It is clear that the initial drive towards edge-computing has been focusing on non-critical applications, but that the momentum and opportunities are likely to lead to increased adoption of edge-computing in CPS, and therefore a need to increasingly deal with multiple attributes of trustworthiness. In the following, we first discuss the findings from our systematic mapping study and then contrast them with the findings of the related surveys. Finally, we discuss the validity of our mapping study.

\subsection*{Discussing the findings}
Among the edge computing solutions, fog computing is the most present 
implementation in the analyzed studies, as seen in Figure \ref{fig:PieChartEdge}. When looking at the distribution within the CPS domains, as in Figure \ref{fig:CPSEdgeTrust}, it can be seen that only in telecom is MEC more frequent than fog computing. This finding is natural given the connection between the telecom industry and MEC. 
A possible explanation for the limited number of papers covering MEC among the CPS domains could be that the telecommunication companies have a strong tradition of patenting (rather than writing papers), note that patents are not covered in this mapping study.
 
By contrast, manufacturing, which is much more focused on fog computing according to the surveyed publications, is also the CPS domain with the highest number of studies about edge-based CPS. 
This is likely because manufacturing technology already has a high degree of automation, sensors, and network capabilities, and so taking the next step to Edge is a relatively small one. Additionally, the search string included "Industry 4.0" which also may favor results within the manufacturing industry. Lastly, the cloudlet architecture is the category with the lowest amount of publications for all of the CPS domains, which could have its explanation in that cloudlets have (so far) mainly intended non-CPS applications.

The consideration of trustworthiness when using edge computing in the context of critical CPS was one of the main motivations for embarking on this study. After analyzing the results in Figure~\ref{fig:PieChartTrust}, it is clear that many of the research efforts in edge-based CPS are only considering trustworthiness of the systems to a very limited extent. In recent years, as shown in Figure \ref{fig:YearTrust} the number of publications related to edge-based CPS has experienced huge growth. Nevertheless, the proportion of those that consider trustworthiness has been constant, around 50\%. Among the ones that consider it, most of them only cover one or two of the attributes of trustworthiness that we have analyzed. 
Regarding those aspects, Predictability and Security received the most attention. For predictability we note a strong interest in various aspects of resource management, which have rendered a classification of "predictability". In hindsight we can say that there are nuances of predictability, essentially referring to "best-effort" (average-case performance) vs. efforts providing some level of guarantees. Thus, not all efforts on predictability are relevant for critical CPS.
Further, safety is the least considered trustworthiness aspect. This is especially noteworthy since many papers indeed refer to various types of critical applications as relevant for edge computing, including for example in manufacturing and transportation (e.g. vehicle platooning). 
The development of safety-critical systems requires adherence to safety standards. 
The current set of standards and thus best practices are not fit for the next generation AI-equipped edge-based CPS, see e.g. \cite{NASA-AssuringSafety, SRA-ECSEL-19, Platforms4CPS2018}. The corresponding challenges are perhaps most prominently seen for automated driving (AD), with the increasing number of efforts that are trying to promote AD safety, see e.g. \cite{SAFAD}. 

Regarding the application type, resource management and real-time application analytics have been the most extensively studied areas, especially using fog computing as the edge implementation, as shown in Figure \ref{fig:PieChartApp}. Power consumption in energy-aware systems, together with task completion latency, are the two most concerned elements during the application. Extensive studies of edge-based CPS, especially regarding computation task offloading \cite{wang2016mobile}, energy-efficient scheduling \cite{yu2018energy}, and resource allocation management \cite{hu2018mobility}, have investigated the natural trade-off between these two factors. The minimization of execution delay and energy consumption in a cooperative edge-based system requires the joint optimization of communication and computation resources between local devices and edge servers. This consideration is usually measured as a weighted-sum function of the delay and energy, and adjusted with different weightings to satisfy the requirements in various use cases. Nevertheless, other factors, such as the operational cost \cite{xu2017online}, network utility \cite{tan2015utility}, quality-of-experience \cite{ning2019deep}, and robustness of transmission network \cite{anwar2017minimax}, are also considered as the objectives to be optimized in the application of edge computing system for CPS.

When analyzing the type of research in Figure \ref{fig:PieChartType}, more than half of the studies have been classified as a solution proposal. When grouping the studies by application type and edge implementation, solution proposal is still the predominant category in almost every group. Evaluation and validation tend to occupy the second and third positions, but there are some research gaps where these two categories are not present, e.g. resource management using cloudlets. As we are embarking on a phase of novel, edge-based CPS in many applications, more effort will be needed on evaluation and validation studies, not the least concerning trustworthiness properties.

Regarding other factors that are influencing the development of edge computing for CPS, this study has analyzed the AI methods used and the inclusion of energy efficiency. Regarding the AI methods, only a third of the studies explicitly state that they use artificial intelligence (Figure~\ref{fig:PieChartAI}). On the other hand, when considering the temporal evolution, (Figure~\ref{fig:YearAI}) one can see an increase in the use of machine learning methods from 2019. Regarding energy efficiency, it is only considered by a rather low fraction of the total number of studies. This value is particularly low in the domains of manufacturing and zero in the healthcare domain.

As CPS are used to integrate new technologies and deployed in settings that span embedded, edge and cloud computing, there are corresponding needs to bridge gaps between the involved research communities. As treated in this paper, this becomes particularly important concerning the edge vs. CPS disciplines. A high-level summary of these findings is illustrated in \autoref{tab:tab4}. Edge computing communities have not had the same exposure to critical applications. Since several trustworthiness attributes have been identified as "research challenges", this would provide a useful starting point for discussions with CPS and dependability fields where these topics have a long tradition. For example, our survey indicates that edge computing research has primarily focused on soft real-time (SRT) systems (where meeting timing requirements is generally not seen as critical), whereas CPS communities have for long studied both hard real-time (HRT) (where missing timing requirements may be critical) and SRT systems, as well as their combination. In any case, as we embark on more open and increasingly complex systems, all trustworthiness attributes face challenges on their own (e.g. for security - more attack surfaces, for safety - learning systems deployed in more open world settings), but also need consideration in conjunction. 

Further, as cyber-physical systems become connected and start to collaborate, this will lead to cyber-physical systems of systems (CPSoS). We believe that many such CPSoS will tend to include edge computing and AI to support many of the coordination and collaboration challenges. Systems of systems are characterized by the operational and managerial independence of their constituent systems, and by emergent behavior, \cite{SoS-Maier}. Take for instance city traffic as a “system”, exemplifying a situation with a multitude of stakeholders, independent evolution (of streets, vehicles, other infrastructure), not always clear responsibilities, and where a change or the introduction of entirely new systems (such as automated vehicles) may cause hard to predict behaviors (emergence).

Finally, although not an aspect we emphasized in our survey, we note that business models are identified as research challenges by the edge computing communities.
To our understanding, CPS has not to the same extent considered this topic and will need to do so as CPS is likely to be increasingly provided as services as part of CPSoS. 


\subsection*{Findings vs. related edge-computing surveys}

It is interesting to reflect on our mapping study findings versus the other state-of-the-art surveys that we summarized in Section~\ref{sec:RelatedWork}, in particular for those covering some flavor of edge-computing.
We note that these survey papers identify a rather broad range of topics as relevant research challenges. Commonly identified challenges include various aspects of resource management, latency, security, privacy, and energy efficiency. Topics identified by a few papers include interoperability-related challenges, governance, business models, architecture, mobility, and application algorithms including data analytics. 

The resource-constrained nature of edge-based systems is highlighted by several papers, where a few call for cost-efficient approaches to security and fault-tolerance. 
The complexity of edge-based systems is touched upon by a few papers directly or indirectly. 
We note that security in itself is a multifaceted topic that would deserve a more in-depth survey. 
While research has addressed or is highlighting selected reliability and availability challenges, security is still mainly identified as a research challenge with an emphasis on privacy and confidentiality -- thus with less coverage of security implications on availability and safety \cite{bakhshi_dependable_2019}. 

In addition, the need to deal with conflicting objectives and multi-objective design is also highlighted by a few surveys (e.g. considering quality-of-service, energy, cost and bandwidth). Specifically, several of the surveys, including \cite{tocze2018, bakhshi_dependable_2019, Yousefpour2019AllOneNeedsKnowFog, Cao2021SurveyEdgeEdgeCloudComputingAssisted}, highlight gaps regarding non-functional properties in terms of trustworthiness/dependability related attributes. As an overall remark, we conclude that the related surveys found similar gaps when it comes to addressing trustworthiness attributes, while there is a lack in considering safety and its relation to other trustworthiness attributes explicitly.
A combined view, drawing upon our findings and the related surveys, is illustrated in \autoref{tab:tab4}.

\begin{table}[]
    \centering
    \caption{High-level overview of findings w.r.t. key CPS properties}
    \label{tab:SummaryFindings}
    \small
    \begin{tabular}{|c|c|c|c|c|}\hline
    \textit{Field vs. Properties} &   \textbf{Safety} &    \textbf{Security} &              \textbf{Predictability} &    \textbf{Energy}              \\\hline
    \textbf{CPS} &                   Yes &       Yes &                   Yes (HRT/SRT) &     Partly              \\\hline
    \textbf{Edge computing} &        No &        Research challenge &    Yes (SRT) &         Research challenge  \\\hline
    \end{tabular}
    \label{tab:tab4}
\end{table}

%


\subsection*{Validity of the results}
Several issues need to be taken into account when conducting a systematic mapping study, which, if unaddressed, can potentially limit the validity of obtained results \cite{Kitchenham07}. 

One such limitation is that this study only considered published papers written in English. For this reason, some relevant contributions in other languages may have been omitted. However, it should be mentioned that this is a limitation with most systematic mapping studies, and the impact is assumed to be small \cite{BOZHINOSKI2019150, abbaspour_asadollah_10_2017}. 
Additionally, since the snowballing process was automated, it is possible that the occasional publication is parsed incorrectly, and thus be considered "unknown" by the automation tool. Such publications would thus not have been properly processed by the management software, and thus omitted from the results. 

Another potential threat to the validity is the subjectivity of the individual researcher during the classification stage. Since only one option is chosen for each category, it can sometimes be hard to assess the core subject matter or the study under review. 
To mitigate this, a validation process was performed where researchers reviewed a randomly sampled subset from the rest of the team. 
No significant discrepancies were identified during this step. Moreover, weekly meetings were performed where all the reviewers participated, to harmonize the concepts and classifications. This process led to several clarifications and in some cases to re-reviews of papers to make sure the same approach was applied to all papers.

Finally, the relevance of the findings may not be representative or relevant if the search string/terms were not appropriate to the corresponding research questions. 
As discussed in this paper, this topic is non-trivial since many concepts and synonyms are used to refer to edge-computing as well as to CPS. We believe that the validity of our mapping study is strengthened with the comparison with the related surveys since they encompass a broader (sometimes slightly different) scope compared to our mapping study. 
For example, the state-of-the-art surveys also include "CPS-" and "edge-" only surveys. 
The performed snowballing also helped to reduce the risk of missing relevant publications.

\section{Recommendations and Future Work}
\label{sec:FutureWork}
As covered by our mapping and the related surveys (Section ~\ref{sec:RelatedWork}), a multitude of topics is already being researched concerning edge computing systems. It is clear that much more research and industrial efforts (including standardization) will be needed in the direction of future edge-based CPS.
We summarize here our analysis of the findings (from the Discussion Section~\ref{sec:Disc}) in terms of recommendations for further research and other efforts that would complement current efforts:
\begin{itemize}
    \item \textit{Further addressing security, safety, and predictability challenges}. Each trustworthiness property needs further research on its own, but also to take the others into account, and moreover, the considerations of multiple simultaneous functional and extra-functional requirements to be considered during design and dealt with during run-time. Research directions include how to deal with security (new vulnerabilities and attack surfaces) and predictability given the dynamics of edge-based CPS (e.g. mobility, and partial failures) and the desire to reason about and tailor latency (e.g. with respect to different quality of services levels) over end-to-end complex computational chains. Edge-computing and communication provide new or enhanced capabilities that "augment CPS", for example by enhancing performance and safety. At the same time, these new capabilities, based on hardware, software and data (with environment dependencies), increase the system complexity and invariably lead to new faults and failure modes, as well as potential unintended effects (emergence) and unintended usage. These effects are likely to introduce new hazards and risks that will require new research to better understand how to systematically deal with risk mitigation and the challenging task of safety assurance/certification in the context of future edge-based CPS, \cite{NASA-AssuringSafety}.
    \item \textit{Addressing the relationships between trustworthiness properties}. This requires an understanding of how these properties relate to each other, can be traded against each-other - ensuring a proper balancing between trustworthiness properties in partly open upgradable systems, and how edge-based CPS can be realized in cost-efficient ways. Important directions here include methodologies for complexity management, run-time reconfiguration, architecture frameworks, and reference architectures. 
    As a common pattern, shared between the trustworthiness attributes considered in this paper, there is a need to investigate how to manage and orchestrate such compute/communication chains to obtain (optimize and trade) the desired properties (e.g. w.r.t. to latency, robustness, availability and so on). Key ingredients here include monitoring, error/anomaly detection, error handling, and ways to deal with system reconfiguration and degrades modes.
    \item \textit{Architecting, platforms and programmability}. Edge-based CPS will involve the (often) dynamic integration of heterogeneous subsystems, with tight internal and external (environment) interactions. These systems often have long life-spans -- over which they are also likely to evolve -- and so must thus be maintainable, upgradeable, debuggable and scalable. Research is needed into platforms and programming models that can enable such properties, along with interoperability, reconfigurability and energy management, while explicitly supporting trustworthiness properties at various levels. We believe that the trustworthiness properties need to be treated as first-class citizens, all the way from reference architectures, over APIs to the programming models. Resilience needs to be provided bottom-up with sufficient tailorability to suite different application needs. Research needs to address new abstractions and architectures in order to find a balance between the increasing complexity (of new mechanisms) and the overall system properties.

    \item \textit{Business models and operational models (contracts) for edge-based CPS}. Edge computing will not only introduce new technology into CPS but in many cases also new stakeholders such as edge computing and communication platform providers and operators. Our findings support a need to further investigate suitable business models and "contracts", that would promote collaborative edge-based CPS, clarify responsibilities and liability. 
    \item \textit{Considering the characteristics and domain-specific requirements of edge-based CPS}. Research and other efforts need to consider the specific characteristics of edge-computing systems in terms of their distributed nature, heterogeneity, dynamics (e.g. potential mobility), resource constraints, and trustworthiness-related requirements. The latter requirements will vary among application domains, 
and in regard to the risks of the domains concerned. 
    \item \textit{Incorporating energy and environmental sustainability considerations into research}. Edge-based CPS form part of an increasingly digitalized society with computing "everywhere". To make this cost-efficient and to minimize environmental impact, circular economy concepts (reuse, repair, re-purpose, etc.) and energy considerations need further research and to be integrated into the overall architecting of future edge-based CPS. As stated in \cite{Hamm.2020}, the sustainable developments generally only receive little attention within the framework of edge computing. Hence, the sustainability should be incorporated in the development of edge computing.
    \item \textit{Emphasis on testbeds and experimental evaluation/validation}. This recommendation follows from the relative novelty of edge-based CPS as a field and the apparently limited emphasis on experimental work. While the limited amount of work could be an indication of early stages, it is important with testbeds for experimentation and learning. This might be even more important for edge-based CPS, as they integrate technologies from telecom, IT/cloud, embedded systems and communications.  
    \item \textit{Forums for networking and collaboration regarding edge-based CPS}. The integration mentioned in the previous bullet(s) requires establishing new forums for interactions between the CPS and edge-computing communities. We also believe that reference architectures and architectural frameworks (first bullet) can help to address the needed cross-domain understanding.
\end{itemize}

As a follow-up to our systematic mapping study, it could be of interest to increase the scope of the study, by incorporating more attributes related to trustworthiness, such as transparency and accountability (\cite{EU.2019}), and also potentially to increase the level of detail by including more related attributes (or sub-attributes), such as resilience, availability, integrity and confidentiality. A further potential direction would be to increase the reach of the study through the inclusion of other search terms, potentially providing further insights. Such directions include incorporating related concepts or characteristic properties with respect to CPS and industry 4.0 like IIoT or dependability. It would also be of interest to provide a more in-depth analysis regarding sustainability and related concepts such as  the  "circular economy". A more fine-grained analysis of the AI methods used for edge-based CPS would also be beneficial. A broader reach covering more of some of the industrial developments could also beneficially be extended to incorporate patents.
Since the whole field of research is growing rapidly, an update to include the newest papers would also beneficial in the next few years. We also believe that nuances of predictability and security could be explored in more detail.

\section{Conclusions}
\label{sec:Conclusions}

The introduction of edge computing for CPS comes as a natural solution to the opportunities at hand, and the current limitations of embedded systems and cloud computing. However, the heterogeneity of "things at the edge" as well as the integration with other fields of computing, has  brought proposals for multiple possible solutions.

This study provides an overview of the current research efforts in the usage of edge computing solutions for critical CPS. Through the analysis and classification of 224 papers, this study provides an overview and insight into the current connections between the two fields and the corresponding research gaps. The analysis motivates a bigger emphasis on research to address trustworthiness-related properties, an aspect that is particularly relevant and necessary for the introduction of critical edge-based CPS. 
\begin{acks}
This research has been carried out as part of the TECoSA Vinnova Competence Center for Trustworthy Edge Computing Systems and Applications at KTH Royal Institute of Technology and in
addition been partly supported through the InSecTT. InSecTT (www.insectt.eu) has received funding from the ECSEL Joint Undertaking (JU) under grant agreement No 876038. The JU receives support from the European Union’s Horizon 2020 research and innovation programme and Austria, Sweden, Spain, Italy, France, Portugal, Ireland, Finland, Slovenia, Poland, Netherlands, Turkey”. The document reflects only the author’s view and the Commission is not responsible for any use that may be made of the information it contains. 
\end{acks}

\bibliographystyle{ACM-Reference-Format}
\bibliography{Bibliography.bib}


\begin{thebibliography}{73}


\ifx \showCODEN    \undefined \def \showCODEN     #1{\unskip}     \fi
\ifx \showDOI      \undefined \def \showDOI       #1{#1}\fi
\ifx \showISBNx    \undefined \def \showISBNx     #1{\unskip}     \fi
\ifx \showISBNxiii \undefined \def \showISBNxiii  #1{\unskip}     \fi
\ifx \showISSN     \undefined \def \showISSN      #1{\unskip}     \fi
\ifx \showLCCN     \undefined \def \showLCCN      #1{\unskip}     \fi
\ifx \shownote     \undefined \def \shownote      #1{#1}          \fi
\ifx \showarticletitle \undefined \def \showarticletitle #1{#1}   \fi
\ifx \showURL      \undefined \def \showURL       {\relax}        \fi
\providecommand\bibfield[2]{#2}
\providecommand\bibinfo[2]{#2}
\providecommand\natexlab[1]{#1}
\providecommand\showeprint[2][]{arXiv:#2}

\bibitem[\protect\citeauthoryear{??}{Ope}{[n.d.]}]%
        {OpenFog.2017}
 \bibinfo{year}{[n.d.]}\natexlab{}.
\newblock \showarticletitle{{IEEE} Standard for Adoption of {OpenFog} Reference
  Architecture for Fog Computing}.
\newblock  (\bibinfo{year}{[n.\,d.]}), \bibinfo{pages}{1--176}.
\newblock
\urldef\tempurl%
\url{https://doi.org/10.1109/IEEESTD.2018.8423800}
\showDOI{\tempurl}
\newblock
\shownote{Conference Name: {IEEE} Std 1934-2018.}


\bibitem[\protect\citeauthoryear{??}{ISO}{2021a}]%
        {ISO-AWI-TS-5083}
 \bibinfo{year}{2021}\natexlab{a}.
\newblock \bibinfo{title}{ISO/AWI TS 5083: Road vehicles — Safety for
  automated driving systems — Design, verification and validation - Under
  development}.
\newblock
\newblock
\urldef\tempurl%
\url{https://www.iso.org/standard/81920.html}
\showURL{%
\tempurl}


\bibitem[\protect\citeauthoryear{??}{ISO}{2021b}]%
        {ISO-SAE-21434}
 \bibinfo{year}{2021}\natexlab{b}.
\newblock \bibinfo{title}{ISO/SAE 21434 - Road vehicles — Cybersecurity
  engineering (under development)}.
\newblock
\newblock


\bibitem[\protect\citeauthoryear{??}{P28}{2021}]%
        {P2846}
 \bibinfo{year}{2021}\natexlab{}.
\newblock \bibinfo{title}{P2846 - Assumptions for Models in Safety-Related
  Automated Vehicle Behavior - Under development}.
\newblock
\newblock
\urldef\tempurl%
\url{https://standards.ieee.org/project/2846.html}
\showURL{%
\tempurl}


\bibitem[\protect\citeauthoryear{{Abbas}, {Zhang}, {Taherkordi}, and
  {Skeie}}{{Abbas} et~al\mbox{.}}{2018}]%
        {Abbas.2018}
\bibfield{author}{\bibinfo{person}{N. {Abbas}}, \bibinfo{person}{Y. {Zhang}},
  \bibinfo{person}{A. {Taherkordi}}, {and} \bibinfo{person}{T. {Skeie}}.}
  \bibinfo{year}{2018}\natexlab{}.
\newblock \showarticletitle{Mobile Edge Computing: A Survey}.
\newblock \bibinfo{journal}{\emph{IEEE Internet of Things Journal}}
  \bibinfo{volume}{5}, \bibinfo{number}{1} (\bibinfo{year}{2018}),
  \bibinfo{pages}{450--465}.
\newblock


\bibitem[\protect\citeauthoryear{Abbaspour~Asadollah, Sundmark, Eldh, Hansson,
  and Afzal}{Abbaspour~Asadollah et~al\mbox{.}}{2017}]%
        {abbaspour_asadollah_10_2017}
\bibfield{author}{\bibinfo{person}{Sara Abbaspour~Asadollah},
  \bibinfo{person}{Daniel Sundmark}, \bibinfo{person}{Sigrid Eldh},
  \bibinfo{person}{Hans Hansson}, {and} \bibinfo{person}{Wasif Afzal}.}
  \bibinfo{year}{2017}\natexlab{}.
\newblock \showarticletitle{10 {Years} of research on debugging concurrent and
  multicore software: a systematic mapping study}.
\newblock \bibinfo{journal}{\emph{Software Quality Journal}}
  \bibinfo{volume}{25}, \bibinfo{number}{1} (\bibinfo{date}{March}
  \bibinfo{year}{2017}), \bibinfo{pages}{49--82}.
\newblock
\showISSN{1573-1367}
\urldef\tempurl%
\url{https://doi.org/10.1007/s11219-015-9301-7}
\showDOI{\tempurl}


\bibitem[\protect\citeauthoryear{{AENEAS}, {ARTEMIS}, and
  associations}{{AENEAS} et~al\mbox{.}}{2019}]%
        {SRA-ECSEL-19}
\bibfield{author}{\bibinfo{person}{{AENEAS}}, \bibinfo{person}{{ARTEMIS}},
  {and} \bibinfo{person}{{EPoSS} associations}.}
  \bibinfo{year}{2019}\natexlab{}.
\newblock \bibinfo{title}{Strategic Research Agenda for Electronic Components
  and Systems}.
\newblock
  \bibinfo{howpublished}{\url{https://www.ecsel.eu/sites/default/files/2019-02/ECS-SRA\%2019\%FINAL.pdf}}.
\newblock


\bibitem[\protect\citeauthoryear{Ahmed, Ahmed, Yaqoob, Shuja, A.~Gani, and
  Shoaib}{Ahmed et~al\mbox{.}}{2017}]%
        {Ahmed2017}
\bibfield{author}{\bibinfo{person}{E. Ahmed}, \bibinfo{person}{A. Ahmed},
  \bibinfo{person}{I. Yaqoob}, \bibinfo{person}{J. Shuja},
  \bibinfo{person}{M.~Imran A.~Gani}, {and} \bibinfo{person}{M. Shoaib}.}
  \bibinfo{year}{2017}\natexlab{}.
\newblock \showarticletitle{Bringing Computation Closer toward the User
  Network: Is Edge Computing the Solution?}
\newblock \bibinfo{journal}{\emph{IEEE Communications Magazine}}
  \bibinfo{volume}{55}, \bibinfo{number}{11} (\bibinfo{year}{2017}),
  \bibinfo{pages}{138--144}.
\newblock


\bibitem[\protect\citeauthoryear{Anwar and Zhu}{Anwar and Zhu}{2017}]%
        {anwar2017minimax}
\bibfield{author}{\bibinfo{person}{Hamza Anwar} {and} \bibinfo{person}{Quanyan
  Zhu}.} \bibinfo{year}{2017}\natexlab{}.
\newblock \showarticletitle{Minimax robust optimal control of multiscale
  linear-quadratic systems}. In \bibinfo{booktitle}{\emph{2017 51st Annual
  Conference on Information Sciences and Systems (CISS)}}. IEEE,
  \bibinfo{pages}{1--6}.
\newblock


\bibitem[\protect\citeauthoryear{{Avizienis}, {Laprie}, {Randell}, and
  {Landwehr}}{{Avizienis} et~al\mbox{.}}{2004}]%
        {Dep}
\bibfield{author}{\bibinfo{person}{A. {Avizienis}}, \bibinfo{person}{J.
  {Laprie}}, \bibinfo{person}{B. {Randell}}, {and} \bibinfo{person}{C.
  {Landwehr}}.} \bibinfo{year}{2004}\natexlab{}.
\newblock \showarticletitle{Basic concepts and taxonomy of dependable and
  secure computing}.
\newblock \bibinfo{journal}{\emph{IEEE Transactions on Dependable and Secure
  Computing}} \bibinfo{volume}{1}, \bibinfo{number}{1} (\bibinfo{year}{2004}),
  \bibinfo{pages}{11--33}.
\newblock


\bibitem[\protect\citeauthoryear{Bakhshi, Rodriguez-Navas, and Hansson}{Bakhshi
  et~al\mbox{.}}{2019}]%
        {bakhshi_dependable_2019}
\bibfield{author}{\bibinfo{person}{Z. Bakhshi}, \bibinfo{person}{G.
  Rodriguez-Navas}, {and} \bibinfo{person}{H. Hansson}.}
  \bibinfo{year}{2019}\natexlab{}.
\newblock \showarticletitle{Dependable {Fog} {Computing}: {A} {Systematic}
  {Literature} {Review}}. In \bibinfo{booktitle}{\emph{2019 45th {Euromicro}
  {Conference} on {Software} {Engineering} and {Advanced} {Applications}
  ({SEAA})}}. \bibinfo{pages}{395--403}.
\newblock
\urldef\tempurl%
\url{https://doi.org/10.1109/SEAA.2019.00066}
\showDOI{\tempurl}


\bibitem[\protect\citeauthoryear{Basir, Qaisar, Ali, Aldwairi, Ashraf, Mahmood,
  and Gidlund}{Basir et~al\mbox{.}}{2019}]%
        {Basir2019FogCE}
\bibfield{author}{\bibinfo{person}{Rabeea Basir}, \bibinfo{person}{Saad~B.
  Qaisar}, \bibinfo{person}{Mudassar Ali}, \bibinfo{person}{Monther Aldwairi},
  \bibinfo{person}{Muhammad~Ikram Ashraf}, \bibinfo{person}{Aamir Mahmood},
  {and} \bibinfo{person}{Mikael Gidlund}.} \bibinfo{year}{2019}\natexlab{}.
\newblock \showarticletitle{Fog Computing Enabling Industrial Internet of
  Things: State-of-the-Art and Research Challenges}.
\newblock \bibinfo{journal}{\emph{Sensors (Basel, Switzerland)}}
  \bibinfo{volume}{19} (\bibinfo{year}{2019}).
\newblock


\bibitem[\protect\citeauthoryear{Bonomi, Milito, Zhu, and Addepalli}{Bonomi
  et~al\mbox{.}}{2012}]%
        {Bonomi.2012}
\bibfield{author}{\bibinfo{person}{Flavio Bonomi}, \bibinfo{person}{Rodolfo
  Milito}, \bibinfo{person}{Jiang Zhu}, {and} \bibinfo{person}{Sateesh
  Addepalli}.} \bibinfo{year}{2012}\natexlab{}.
\newblock \showarticletitle{Fog Computing and Its Role in the Internet of
  Things}. In \bibinfo{booktitle}{\emph{Proceedings of the First Edition of the
  MCC Workshop on Mobile Cloud Computing}} (Helsinki, Finland)
  \emph{(\bibinfo{series}{MCC ’12})}. \bibinfo{publisher}{Association for
  Computing Machinery}, \bibinfo{address}{New York, NY, USA},
  \bibinfo{pages}{13–16}.
\newblock
\showISBNx{9781450315197}
\urldef\tempurl%
\url{https://doi.org/10.1145/2342509.2342513}
\showDOI{\tempurl}


\bibitem[\protect\citeauthoryear{Bozhinoski, Ruscio], Malavolta, Pelliccione,
  and Crnkovic}{Bozhinoski et~al\mbox{.}}{2019}]%
        {BOZHINOSKI2019150}
\bibfield{author}{\bibinfo{person}{Darko Bozhinoski},
  \bibinfo{person}{Davide~[Di Ruscio]}, \bibinfo{person}{Ivano Malavolta},
  \bibinfo{person}{Patrizio Pelliccione}, {and} \bibinfo{person}{Ivica
  Crnkovic}.} \bibinfo{year}{2019}\natexlab{}.
\newblock \showarticletitle{Safety for mobile robotic systems: A systematic
  mapping study from a software engineering perspective}.
\newblock \bibinfo{journal}{\emph{Journal of Systems and Software}}
  \bibinfo{volume}{151} (\bibinfo{year}{2019}), \bibinfo{pages}{150 -- 179}.
\newblock
\showISSN{0164-1212}
\urldef\tempurl%
\url{https://doi.org/10.1016/j.jss.2019.02.021}
\showDOI{\tempurl}


\bibitem[\protect\citeauthoryear{Caiza, Saeteros, Oñate, and Garcia}{Caiza
  et~al\mbox{.}}{2020}]%
        {CAIZA2020e03706}
\bibfield{author}{\bibinfo{person}{Gustavo Caiza}, \bibinfo{person}{Morelva
  Saeteros}, \bibinfo{person}{William Oñate}, {and}
  \bibinfo{person}{Marcelo~V. Garcia}.} \bibinfo{year}{2020}\natexlab{}.
\newblock \showarticletitle{Fog computing at industrial level, architecture,
  latency, energy, and security: A review}.
\newblock \bibinfo{journal}{\emph{Heliyon}} \bibinfo{volume}{6},
  \bibinfo{number}{4} (\bibinfo{year}{2020}), \bibinfo{pages}{e03706}.
\newblock
\showISSN{2405-8440}
\urldef\tempurl%
\url{https://doi.org/10.1016/j.heliyon.2020.e03706}
\showDOI{\tempurl}


\bibitem[\protect\citeauthoryear{Cao, Hu, Shi, Colombo, Karnouskos, and Li}{Cao
  et~al\mbox{.}}{2021}]%
        {Cao2021SurveyEdgeEdgeCloudComputingAssisted}
\bibfield{author}{\bibinfo{person}{Kun Cao}, \bibinfo{person}{Shiyan Hu},
  \bibinfo{person}{Yang Shi}, \bibinfo{person}{Armando~Walter Colombo},
  \bibinfo{person}{Stamatis Karnouskos}, {and} \bibinfo{person}{Xin Li}.}
  \bibinfo{year}{2021}\natexlab{}.
\newblock \showarticletitle{A {{Survey}} on {{Edge}} and {{Edge}}-{{Cloud
  Computing Assisted Cyber}}-{{Physical Systems}}}.
\newblock \bibinfo{journal}{\emph{IEEE Transactions on Industrial Informatics}}
  \bibinfo{volume}{17}, \bibinfo{number}{11} (\bibinfo{date}{Nov.}
  \bibinfo{year}{2021}), \bibinfo{pages}{7806--7819}.
\newblock
\showISSN{1941-0050}
\urldef\tempurl%
\url{https://doi.org/10.1109/TII.2021.3073066}
\showDOI{\tempurl}


\bibitem[\protect\citeauthoryear{Center}{Center}{[n.d.]}]%
        {NistGlossary}
\bibfield{author}{\bibinfo{person}{NIST Computer Security~Resource Center}.}
  \bibinfo{year}{[n.d.]}\natexlab{}.
\newblock \bibinfo{title}{Glossary}.
\newblock
  \bibinfo{howpublished}{\url{https://csrc.nist.gov/glossary/term/trustworthiness}}.
\newblock


\bibitem[\protect\citeauthoryear{Chen}{Chen}{2017}]%
        {Chen-2017}
\bibfield{author}{\bibinfo{person}{Hong Chen}.}
  \bibinfo{year}{2017}\natexlab{}.
\newblock \showarticletitle{Applications of Cyber-Physical System: A Literature
  Review}.
\newblock \bibinfo{journal}{\emph{Journal of Industrial Integration and
  Management - Special Issue: Cyber Physical Systems in Industrial
  Integration}} \bibinfo{volume}{2}, \bibinfo{number}{3}
  (\bibinfo{year}{2017}).
\newblock
\urldef\tempurl%
\url{https://doi.org/10.1142/S2424862217500129}
\showDOI{\tempurl}


\bibitem[\protect\citeauthoryear{{Cintuglu}, {Mohammed}, {Akkaya}, and
  {Uluagac}}{{Cintuglu} et~al\mbox{.}}{2017}]%
        {7740849}
\bibfield{author}{\bibinfo{person}{M.~H. {Cintuglu}}, \bibinfo{person}{O.~A.
  {Mohammed}}, \bibinfo{person}{K. {Akkaya}}, {and} \bibinfo{person}{A.~S.
  {Uluagac}}.} \bibinfo{year}{2017}\natexlab{}.
\newblock \showarticletitle{A Survey on Smart Grid Cyber-Physical System
  Testbeds}.
\newblock \bibinfo{journal}{\emph{IEEE Communications Surveys Tutorials}}
  \bibinfo{volume}{19}, \bibinfo{number}{1} (\bibinfo{year}{2017}),
  \bibinfo{pages}{446--464}.
\newblock


\bibitem[\protect\citeauthoryear{Commission}{Commission}{2010}]%
        {IEC}
\bibfield{author}{\bibinfo{person}{International~Electrotechnical Commission}.}
  \bibinfo{year}{2010}\natexlab{}.
\newblock \bibinfo{title}{Functional Safety of E/E/PE Safety-Related Systems,
  Standard 61508:2010}.
\newblock
\newblock


\bibitem[\protect\citeauthoryear{Consortium}{Consortium}{2017}]%
        {ECC}
\bibfield{author}{\bibinfo{person}{Edge~Computing Consortium}.}
  \bibinfo{year}{2017}\natexlab{}.
\newblock \bibinfo{title}{Edge Computing Reference Architecture 2.0}.
\newblock
  \bibinfo{howpublished}{\url{http://en.ecconsortium.net/Uploads/file/20180328/1522232376480704.pdf}}.
\newblock


\bibitem[\protect\citeauthoryear{Consortium}{Consortium}{2018}]%
        {IIC}
\bibfield{author}{\bibinfo{person}{Industrial~Internet Consortium}.}
  \bibinfo{year}{2018}\natexlab{}.
\newblock \bibinfo{title}{Edge Computing Reference Architecture 2.0}.
\newblock \bibinfo{howpublished}{\url{https://www.iiconsortium.org/IIRA.htm}}.
\newblock


\bibitem[\protect\citeauthoryear{Cyber-Physical Systems Public Working~Group
  and Cyber-Physical Systems Program~Office}{Cyber-Physical Systems Public
  Working~Group and Cyber-Physical Systems Program~Office}{2017}]%
        {NIST-CPS}
\bibfield{author}{\bibinfo{person}{Smart~Grid Cyber-Physical Systems Public
  Working~Group} {and} \bibinfo{person}{Engineering~Laboratory Cyber-Physical
  Systems Program~Office}.} \bibinfo{year}{2017}\natexlab{}.
\newblock \bibinfo{title}{Framework for Cyber-Physical Systems: Volume 1,
  Overview, Version 1.0, NIST Spec. Publ. 1500-201}.
\newblock
\newblock
\urldef\tempurl%
\url{https://doi.org/10.6028/NIST.SP.1500-201}
\showDOI{\tempurl}


\bibitem[\protect\citeauthoryear{Damm, Sztipanovits, Baras, Beetz, Bensalem,
  Broy, Grosu, Krogh, Lee, Ruess, Sangiovanni-Vincentelli, and Sifakis}{Damm
  et~al\mbox{.}}{2016}]%
        {damm16}
\bibfield{author}{\bibinfo{person}{Werner Damm}, \bibinfo{person}{Janos
  Sztipanovits}, \bibinfo{person}{John~S. Baras}, \bibinfo{person}{Klaus
  Beetz}, \bibinfo{person}{Saddek Bensalem}, \bibinfo{person}{Manfred Broy},
  \bibinfo{person}{Radu Grosu}, \bibinfo{person}{Bruce~H. Krogh},
  \bibinfo{person}{Insup Lee}, \bibinfo{person}{Harald Ruess},
  \bibinfo{person}{Alberto~L. Sangiovanni-Vincentelli}, {and}
  \bibinfo{person}{Joseph Sifakis}.} \bibinfo{year}{2016}\natexlab{}.
\newblock \bibinfo{title}{Towards a Cross-Cutting Science of Cyber-Physical
  Systems for Mastering all-Important Engineering Challenges}.
\newblock \bibinfo{howpublished}{\url{https://cps-vo.org/node/27006}}.
\newblock


\bibitem[\protect\citeauthoryear{{Dolui} and {Datta}}{{Dolui} and
  {Datta}}{2017}]%
        {8016213}
\bibfield{author}{\bibinfo{person}{K. {Dolui}} {and} \bibinfo{person}{S.~K.
  {Datta}}.} \bibinfo{year}{2017}\natexlab{}.
\newblock \showarticletitle{Comparison of edge computing implementations: Fog
  computing, cloudlet and mobile edge computing}. In
  \bibinfo{booktitle}{\emph{2017 Global Internet of Things Summit (GIoTS)}}.
  \bibinfo{pages}{1--6}.
\newblock


\bibitem[\protect\citeauthoryear{Elazhary}{Elazhary}{2019}]%
        {ELAZHARY2019105}
\bibfield{author}{\bibinfo{person}{Hanan Elazhary}.}
  \bibinfo{year}{2019}\natexlab{}.
\newblock \showarticletitle{Internet of Things (IoT), mobile cloud, cloudlet,
  mobile IoT, IoT cloud, fog, mobile edge, and edge emerging computing
  paradigms: Disambiguation and research directions}.
\newblock \bibinfo{journal}{\emph{Journal of Network and Computer
  Applications}}  \bibinfo{volume}{128} (\bibinfo{year}{2019}),
  \bibinfo{pages}{105 -- 140}.
\newblock
\showISSN{1084-8045}
\urldef\tempurl%
\url{https://doi.org/10.1016/j.jnca.2018.10.021}
\showDOI{\tempurl}


\bibitem[\protect\citeauthoryear{et~al.}{et~al.}{2018}]%
        {NASA-AssuringSafety}
\bibfield{author}{\bibinfo{person}{E.~Alves et al.}}
  \bibinfo{year}{2018}\natexlab{}.
\newblock \bibinfo{title}{Considerations in Assuring Safety of Increasingly
  Autonomous Systems}.
\newblock
  \bibinfo{howpublished}{\url{https://ntrs.nasa.gov/search.jsp?R=20180006312}}.
\newblock


\bibitem[\protect\citeauthoryear{et~al.}{et~al.}{2019}]%
        {SAFAD}
\bibfield{author}{\bibinfo{person}{Matthew~Wood et al.}}
  \bibinfo{year}{2019}\natexlab{}.
\newblock \bibinfo{title}{Safety First for Automated Driving}.
\newblock
  \bibinfo{howpublished}{\url{https://www.daimler.com/documents/innovation/other/safety-first-for-automated-driving.pdf}}.
\newblock


\bibitem[\protect\citeauthoryear{Firesmith}{Firesmith}{2003}]%
        {Firesmith}
\bibfield{author}{\bibinfo{person}{D.~G. Firesmith}.}
  \bibinfo{year}{2003}\natexlab{}.
\newblock \bibinfo{booktitle}{\emph{Common concepts underlying safety,
  security, and survivability engineering}}.
\newblock \bibinfo{type}{{T}echnical {R}eport} Technical Note:
  CMU/SEI-2003-TN-033. \bibinfo{institution}{CMU/SEI}.
\newblock


\bibitem[\protect\citeauthoryear{{Gonzalez}, {Goya}, {de Fatima Pereira},
  {Langona}, {Silva}, {Melo de Brito Carvalho}, {Miers}, {Mångs}, and
  {Sefidcon}}{{Gonzalez} et~al\mbox{.}}{2016}]%
        {Gonzalez2016}
\bibfield{author}{\bibinfo{person}{N.~M. {Gonzalez}}, \bibinfo{person}{W.~A.
  {Goya}}, \bibinfo{person}{R. {de Fatima Pereira}}, \bibinfo{person}{K.
  {Langona}}, \bibinfo{person}{E.~A. {Silva}}, \bibinfo{person}{T.~C. {Melo de
  Brito Carvalho}}, \bibinfo{person}{C.~C. {Miers}}, \bibinfo{person}{J.
  {Mångs}}, {and} \bibinfo{person}{A. {Sefidcon}}.}
  \bibinfo{year}{2016}\natexlab{}.
\newblock \showarticletitle{Fog computing: Data analytics and cloud distributed
  processing on the network edges}. In \bibinfo{booktitle}{\emph{2016 35th
  International Conference of the Chilean Computer Science Society (SCCC)}}.
  \bibinfo{pages}{1--9}.
\newblock


\bibitem[\protect\citeauthoryear{Griffor}{Griffor}{2017}]%
        {SecHandbook}
\bibfield{editor}{\bibinfo{person}{Edward Griffor}} (Ed.).
  \bibinfo{year}{2017}\natexlab{}.
\newblock \bibinfo{booktitle}{\emph{Handbook of System Safety and Security}}.
\newblock \bibinfo{publisher}{Syngress}.
\newblock
\showISBNx{978-0-12-803773-7}
\urldef\tempurl%
\url{https://doi.org/10.1016/C2014-0-05033-2}
\showDOI{\tempurl}


\bibitem[\protect\citeauthoryear{H2020}{H2020}{2017}]%
        {FAR}
\bibfield{author}{\bibinfo{person}{FAR-EDGE~Project H2020}.}
  \bibinfo{year}{2017}\natexlab{}.
\newblock \bibinfo{title}{FAR-EDGE Architecture and Components Specification}.
\newblock
  \bibinfo{howpublished}{\url{https://ec.europa.eu/research/participants/documents/downloadPublic?documentIds=080166e5b3996c23&appId=PPGMS}}.
\newblock


\bibitem[\protect\citeauthoryear{Hamm, Willner, and Schieferdecker}{Hamm
  et~al\mbox{.}}{2020}]%
        {Hamm.2020}
\bibfield{author}{\bibinfo{person}{Andrea Hamm}, \bibinfo{person}{Alexander
  Willner}, {and} \bibinfo{person}{Ina Schieferdecker}.}
  \bibinfo{year}{2020}\natexlab{}.
\newblock \showarticletitle{Edge Computing: A Comprehensive Survey of Current
  Initiatives and a Roadmap for a Sustainable Edge Computing Development}. In
  \bibinfo{booktitle}{\emph{Proceedings of the 15th International Conference on
  Wirtschaftsinformatik (2020)}}. \bibinfo{publisher}{GITO Verlag},
  \bibinfo{pages}{694–709}.
\newblock


\bibitem[\protect\citeauthoryear{Hancock, Billings, Schaefer, Chen, de~Visser,
  and Parasuraman}{Hancock et~al\mbox{.}}{2011}]%
        {Hanckock-Security}
\bibfield{author}{\bibinfo{person}{Peter~A. Hancock},
  \bibinfo{person}{Deborah~R. Billings}, \bibinfo{person}{Kristin~E. Schaefer},
  \bibinfo{person}{Jessie Y.~C. Chen}, \bibinfo{person}{Ewart~J. de Visser},
  {and} \bibinfo{person}{Raja Parasuraman}.} \bibinfo{year}{2011}\natexlab{}.
\newblock \showarticletitle{A Meta-Analysis of Factors Affecting Trust in
  Human-Robot Interaction}.
\newblock \bibinfo{journal}{\emph{Human Factors}} \bibinfo{volume}{53},
  \bibinfo{number}{5} (\bibinfo{year}{2011}), \bibinfo{pages}{517--527}.
\newblock
\urldef\tempurl%
\url{https://doi.org/10.1177/0018720811417254}
\showDOI{\tempurl}
\showeprint{https://doi.org/10.1177/0018720811417254}
\newblock
\shownote{PMID: 22046724.}


\bibitem[\protect\citeauthoryear{{Hao}, {Novak}, {Yi}, and {Li}}{{Hao}
  et~al\mbox{.}}{2017}]%
        {7867731}
\bibfield{author}{\bibinfo{person}{Z. {Hao}}, \bibinfo{person}{E. {Novak}},
  \bibinfo{person}{S. {Yi}}, {and} \bibinfo{person}{Q. {Li}}.}
  \bibinfo{year}{2017}\natexlab{}.
\newblock \showarticletitle{Challenges and Software Architecture for Fog
  Computing}.
\newblock \bibinfo{journal}{\emph{IEEE Internet Computing}}
  \bibinfo{volume}{21}, \bibinfo{number}{2} (\bibinfo{year}{2017}),
  \bibinfo{pages}{44--53}.
\newblock


\bibitem[\protect\citeauthoryear{{High-Level Expert Group on AI of European
  Commission}}{{High-Level Expert Group on AI of European Commission}}{2019}]%
        {EU.2019}
\bibfield{author}{\bibinfo{person}{{High-Level Expert Group on AI of European
  Commission}}.} \bibinfo{year}{2019}\natexlab{}.
\newblock \bibinfo{booktitle}{\emph{Ethics guidelines for trustworthy AI}}.
\newblock Brussels.
\newblock
\urldef\tempurl%
\url{https://ec.europa.eu/digital-single-market/en/news/ethics-guidelines-trustworthy-ai}
\showURL{%
\tempurl}


\bibitem[\protect\citeauthoryear{{High-Level Expert Group on AI of European
  Commission (AI HLEG}}{{High-Level Expert Group on AI of European Commission
  (AI HLEG}}{2021}]%
        {EU.2021.web}
\bibfield{author}{\bibinfo{person}{{High-Level Expert Group on AI of European
  Commission (AI HLEG}}.} \bibinfo{year}{2021}\natexlab{}.
\newblock \bibinfo{booktitle}{\emph{Overview of deliverables from the AI
  HLEG}}.
\newblock
\urldef\tempurl%
\url{https://digital-strategy.ec.europa.eu/en/policies/expert-group-ai}
\showURL{%
\tempurl}


\bibitem[\protect\citeauthoryear{Hu et~al\mbox{.}}{Hu et~al\mbox{.}}{2018}]%
        {hu2018mobility}
\bibfield{author}{\bibinfo{person}{Rose~Qingyang Hu} {et~al\mbox{.}}}
  \bibinfo{year}{2018}\natexlab{}.
\newblock \showarticletitle{Mobility-aware edge caching and computing in
  vehicle networks: A deep reinforcement learning}.
\newblock \bibinfo{journal}{\emph{IEEE Transactions on Vehicular Technology}}
  \bibinfo{volume}{67}, \bibinfo{number}{11} (\bibinfo{year}{2018}),
  \bibinfo{pages}{10190--10203}.
\newblock


\bibitem[\protect\citeauthoryear{Intel-SAP}{Intel-SAP}{2018}]%
        {intel}
\bibfield{author}{\bibinfo{person}{Intel-SAP}.}
  \bibinfo{year}{2018}\natexlab{}.
\newblock \bibinfo{title}{IoT Joint Reference Architecture from Intel and SAP}.
\newblock
  \bibinfo{howpublished}{\url{https://www.intel.com/content/dam/www/public/us/en/documents/reference-architectures/sapiot-reference-architecture.pdf}}.
\newblock


\bibitem[\protect\citeauthoryear{{International Organization for
  Standardization}}{{International Organization for Standardization}}{2018}]%
        {ISO26262}
\bibfield{author}{\bibinfo{person}{{International Organization for
  Standardization}}.} \bibinfo{year}{2018}\natexlab{}.
\newblock \bibinfo{title}{{ISO 26262: Road vehicles--Functional safety}}.
\newblock
\newblock
\urldef\tempurl%
\url{https://www.iso.org/obp/ui/{\#}iso:std:iso:26262:-1:en}
\showURL{%
\tempurl}


\bibitem[\protect\citeauthoryear{{International Organization for
  Standardization}}{{International Organization for Standardization}}{2019}]%
        {SOTIF}
\bibfield{author}{\bibinfo{person}{{International Organization for
  Standardization}}.} \bibinfo{year}{2019}\natexlab{}.
\newblock \bibinfo{booktitle}{\emph{{Road vehicles - Safety of the intended
  functionality (ISO 21448)}}}.
\newblock \bibinfo{type}{{T}echnical {R}eport}.
\newblock
\showISBNx{978 0 580 94502 1}
\urldef\tempurl%
\url{https://www.iso.org/standard/70939.html}
\showURL{%
\tempurl}


\bibitem[\protect\citeauthoryear{Khan, Ahmed, Hakak, Yaqoob, and Ahmed}{Khan
  et~al\mbox{.}}{2019}]%
        {Khan2019}
\bibfield{author}{\bibinfo{person}{Wazir~Zada Khan}, \bibinfo{person}{Ejaz
  Ahmed}, \bibinfo{person}{Saqib Hakak}, \bibinfo{person}{Ibrar Yaqoob}, {and}
  \bibinfo{person}{Arif Ahmed}.} \bibinfo{year}{2019}\natexlab{}.
\newblock \showarticletitle{Edge computing: A survey}.
\newblock \bibinfo{journal}{\emph{In Future Generation Computer Systems}}
  \bibinfo{volume}{97} (\bibinfo{date}{8} \bibinfo{year}{2019}),
  \bibinfo{pages}{219--235}.
\newblock
\urldef\tempurl%
\url{https://doi.org/10.1016/j.future.2019.02.050}
\showDOI{\tempurl}


\bibitem[\protect\citeauthoryear{Kitchenham and Charters}{Kitchenham and
  Charters}{2007}]%
        {Kitchenham07}
\bibfield{author}{\bibinfo{person}{B. Kitchenham} {and} \bibinfo{person}{S
  Charters}.} \bibinfo{year}{2007}\natexlab{}.
\newblock \bibinfo{title}{Guidelines for performing Systematic Literature
  Reviews in Software Engineering}.
\newblock
\newblock


\bibitem[\protect\citeauthoryear{Lu}{Lu}{2017}]%
        {Lu-2017}
\bibfield{author}{\bibinfo{person}{Yang Lu}.} \bibinfo{year}{2017}\natexlab{}.
\newblock \showarticletitle{Cyber Physical System (CPS)-Based Industry 4.0: A
  Survey}.
\newblock \bibinfo{journal}{\emph{Journal of Industrial Integration and
  Management - Special Issue: Cyber Physical Systems in Industrial
  Integration}} \bibinfo{volume}{2}, \bibinfo{number}{3}
  (\bibinfo{year}{2017}).
\newblock
\urldef\tempurl%
\url{https://doi.org/10.1142/S2424862217500142}
\showDOI{\tempurl}


\bibitem[\protect\citeauthoryear{Maier}{Maier}{1998}]%
        {SoS-Maier}
\bibfield{author}{\bibinfo{person}{Mark~W. Maier}.}
  \bibinfo{year}{1998}\natexlab{}.
\newblock \showarticletitle{Architecting principles for systems-of-systems}.
\newblock \bibinfo{journal}{\emph{Systems Engineering}} \bibinfo{volume}{1},
  \bibinfo{number}{4} (\bibinfo{year}{1998}), \bibinfo{pages}{267--284}.
\newblock
\urldef\tempurl%
\url{https://doi.org/10.1002/(SICI)1520-6858(1998)1:4<267::AID-SYS3>3.0.CO;2-D}
\showDOI{\tempurl}


\bibitem[\protect\citeauthoryear{{Mao}, {You}, {Zhang}, {Huang}, and
  {Letaief}}{{Mao} et~al\mbox{.}}{2017}]%
        {Mao2017}
\bibfield{author}{\bibinfo{person}{Y. {Mao}}, \bibinfo{person}{C. {You}},
  \bibinfo{person}{J. {Zhang}}, \bibinfo{person}{K. {Huang}}, {and}
  \bibinfo{person}{K.~B. {Letaief}}.} \bibinfo{year}{2017}\natexlab{}.
\newblock \showarticletitle{A Survey on Mobile Edge Computing: The
  Communication Perspective}.
\newblock \bibinfo{journal}{\emph{IEEE Communications Surveys Tutorials}}
  \bibinfo{volume}{19}, \bibinfo{number}{4} (\bibinfo{year}{2017}),
  \bibinfo{pages}{2322--2358}.
\newblock


\bibitem[\protect\citeauthoryear{{Mouradian}, {Naboulsi}, {Yangui}, {Glitho},
  {Morrow}, and {Polakos}}{{Mouradian} et~al\mbox{.}}{2018}]%
        {8100873}
\bibfield{author}{\bibinfo{person}{C. {Mouradian}}, \bibinfo{person}{D.
  {Naboulsi}}, \bibinfo{person}{S. {Yangui}}, \bibinfo{person}{R.~H. {Glitho}},
  \bibinfo{person}{M.~J. {Morrow}}, {and} \bibinfo{person}{P.~A. {Polakos}}.}
  \bibinfo{year}{2018}\natexlab{}.
\newblock \showarticletitle{A Comprehensive Survey on Fog Computing:
  State-of-the-Art and Research Challenges}.
\newblock \bibinfo{journal}{\emph{IEEE Communications Surveys Tutorials}}
  \bibinfo{volume}{20}, \bibinfo{number}{1} (\bibinfo{year}{2018}),
  \bibinfo{pages}{416--464}.
\newblock


\bibitem[\protect\citeauthoryear{{National Institute of Standards and
  Technology Cyber Physical Systems Public Working Group}}{{National Institute
  of Standards and Technology Cyber Physical Systems Public Working
  Group}}{2016}]%
        {nist16}
\bibfield{author}{\bibinfo{person}{{National Institute of Standards and
  Technology Cyber Physical Systems Public Working Group}}.}
  \bibinfo{year}{2016}\natexlab{}.
\newblock \bibinfo{title}{Framework for Cyber-Physical Systems - Release 1.0}.
\newblock \bibinfo{howpublished}{\url{https://pages.nist.gov/cpspwg/}}.
\newblock


\bibitem[\protect\citeauthoryear{Ning, Dong, Wang, Rodrigues, and Xia}{Ning
  et~al\mbox{.}}{2019}]%
        {ning2019deep}
\bibfield{author}{\bibinfo{person}{Zhaolong Ning}, \bibinfo{person}{Peiran
  Dong}, \bibinfo{person}{Xiaojie Wang}, \bibinfo{person}{Joel~JPC Rodrigues},
  {and} \bibinfo{person}{Feng Xia}.} \bibinfo{year}{2019}\natexlab{}.
\newblock \showarticletitle{Deep reinforcement learning for vehicular edge
  computing: An intelligent offloading system}.
\newblock \bibinfo{journal}{\emph{ACM Transactions on Intelligent Systems and
  Technology (TIST)}} \bibinfo{volume}{10}, \bibinfo{number}{6}
  (\bibinfo{year}{2019}), \bibinfo{pages}{1--24}.
\newblock


\bibitem[\protect\citeauthoryear{of~Technology}{of~Technology}{[n.d.]}]%
        {TECoSA}
\bibfield{author}{\bibinfo{person}{KTH Royal~Institute of Technology}.}
  \bibinfo{year}{[n.d.]}\natexlab{}.
\newblock \bibinfo{title}{TECoSA research center on Trustworthy Edge Computing
  Systems and Applications}.
\newblock
\newblock
\urldef\tempurl%
\url{https://www.tecosa.center.kth.se/ (accessed July 2021)}
\showURL{%
\tempurl}


\bibitem[\protect\citeauthoryear{Perera, Qin, Estrella, Reiff-Marganiec, and
  Vasilakos}{Perera et~al\mbox{.}}{2017}]%
        {10.1145/3057266}
\bibfield{author}{\bibinfo{person}{Charith Perera}, \bibinfo{person}{Yongrui
  Qin}, \bibinfo{person}{Julio~C. Estrella}, \bibinfo{person}{Stephan
  Reiff-Marganiec}, {and} \bibinfo{person}{Athanasios~V. Vasilakos}.}
  \bibinfo{year}{2017}\natexlab{}.
\newblock \showarticletitle{Fog Computing for Sustainable Smart Cities: A
  Survey}.
\newblock \bibinfo{journal}{\emph{ACM Comput. Surv.}} \bibinfo{volume}{50},
  \bibinfo{number}{3}, Article \bibinfo{articleno}{32} (\bibinfo{date}{June}
  \bibinfo{year}{2017}), \bibinfo{numpages}{43}~pages.
\newblock
\showISSN{0360-0300}
\urldef\tempurl%
\url{https://doi.org/10.1145/3057266}
\showDOI{\tempurl}


\bibitem[\protect\citeauthoryear{Petersen, Feldt, Mujtaba, and
  Mattsson}{Petersen et~al\mbox{.}}{2008}]%
        {PETERSEN2008}
\bibfield{author}{\bibinfo{person}{Kai Petersen}, \bibinfo{person}{Robert
  Feldt}, \bibinfo{person}{Shahid Mujtaba}, {and} \bibinfo{person}{Michael
  Mattsson}.} \bibinfo{year}{2008}\natexlab{}.
\newblock \showarticletitle{Systematic Mapping Studies in Software
  Engineering}. In \bibinfo{booktitle}{\emph{Proceedings of the 12th
  International Conference on Evaluation and Assessment in Software
  Engineering}} (Italy) \emph{(\bibinfo{series}{EASE'08})}.
  \bibinfo{publisher}{BCS Learning Development Ltd.},
  \bibinfo{address}{Swindon, GBR}, \bibinfo{pages}{68–77}.
\newblock


\bibitem[\protect\citeauthoryear{Petersen, Vakkalanka, and Kuzniarz}{Petersen
  et~al\mbox{.}}{2015}]%
        {PETERSEN20151}
\bibfield{author}{\bibinfo{person}{Kai Petersen}, \bibinfo{person}{Sairam
  Vakkalanka}, {and} \bibinfo{person}{Ludwik Kuzniarz}.}
  \bibinfo{year}{2015}\natexlab{}.
\newblock \showarticletitle{Guidelines for conducting systematic mapping
  studies in software engineering: An update}.
\newblock \bibinfo{journal}{\emph{Information and Software Technology}}
  \bibinfo{volume}{64} (\bibinfo{year}{2015}), \bibinfo{pages}{1 -- 18}.
\newblock


\bibitem[\protect\citeauthoryear{Rajkumar, Lee, Sha, and Stankovic}{Rajkumar
  et~al\mbox{.}}{2010}]%
        {raj10}
\bibfield{author}{\bibinfo{person}{Ragunathan Rajkumar}, \bibinfo{person}{Insup
  Lee}, \bibinfo{person}{Lui Sha}, {and} \bibinfo{person}{John Stankovic}.}
  \bibinfo{year}{2010}\natexlab{}.
\newblock \showarticletitle{Cyber-physical systems: The next computing
  revolution}. In \bibinfo{booktitle}{\emph{Proc. 47th Design Automation
  Conference}}. \bibinfo{publisher}{IEEE}, \bibinfo{address}{Anaheim, CA, USA}.
\newblock
\urldef\tempurl%
\url{https://doi.org/10.1145/1837274.1837461}
\showDOI{\tempurl}


\bibitem[\protect\citeauthoryear{Ren, Zhang, He, Zhang, and Li}{Ren
  et~al\mbox{.}}{2019}]%
        {renSurveyEndEdgeCloudOrchestrated2019}
\bibfield{author}{\bibinfo{person}{Ju Ren}, \bibinfo{person}{Deyu Zhang},
  \bibinfo{person}{Shiwen He}, \bibinfo{person}{Yaoxue Zhang}, {and}
  \bibinfo{person}{Tao Li}.} \bibinfo{year}{2019}\natexlab{}.
\newblock \showarticletitle{A {{Survey}} on {{End}}-{{Edge}}-{{Cloud
  Orchestrated Network Computing Paradigms}}: {{Transparent Computing}},
  {{Mobile Edge Computing}}, {{Fog Computing}}, and {{Cloudlet}}}.
\newblock \bibinfo{journal}{\emph{Comput. Surveys}} \bibinfo{volume}{52},
  \bibinfo{number}{6} (\bibinfo{date}{Oct.} \bibinfo{year}{2019}),
  \bibinfo{pages}{125:1--125:36}.
\newblock
\showISSN{0360-0300}
\urldef\tempurl%
\url{https://doi.org/10.1145/3362031}
\showDOI{\tempurl}


\bibitem[\protect\citeauthoryear{Satyanarayanan}{Satyanarayanan}{2017}]%
        {sat17}
\bibfield{author}{\bibinfo{person}{M. Satyanarayanan}.}
  \bibinfo{year}{2017}\natexlab{}.
\newblock \showarticletitle{The Emergence of Edge Computing}.
\newblock \bibinfo{journal}{\emph{IEEE Computer}} \bibinfo{volume}{50},
  \bibinfo{number}{1} (\bibinfo{year}{2017}).
\newblock


\bibitem[\protect\citeauthoryear{{Satyanarayanan}, {Bahl}, {Caceres}, and
  {Davies}}{{Satyanarayanan} et~al\mbox{.}}{2009}]%
        {Satyanarayanan.2009}
\bibfield{author}{\bibinfo{person}{M. {Satyanarayanan}}, \bibinfo{person}{P.
  {Bahl}}, \bibinfo{person}{R. {Caceres}}, {and} \bibinfo{person}{N.
  {Davies}}.} \bibinfo{year}{2009}\natexlab{}.
\newblock \showarticletitle{The Case for VM-Based Cloudlets in Mobile
  Computing}.
\newblock \bibinfo{journal}{\emph{IEEE Pervasive Computing}}
  \bibinfo{volume}{8}, \bibinfo{number}{4} (\bibinfo{year}{2009}),
  \bibinfo{pages}{14--23}.
\newblock


\bibitem[\protect\citeauthoryear{Sittón-Candanedo, Alonso, Corchado,
  Rodríguez-González, and Casado-Vara}{Sittón-Candanedo
  et~al\mbox{.}}{2019}]%
        {Sitton.2019}
\bibfield{author}{\bibinfo{person}{Inés Sittón-Candanedo},
  \bibinfo{person}{Ricardo~S. Alonso}, \bibinfo{person}{Juan~M. Corchado},
  \bibinfo{person}{Sara Rodríguez-González}, {and} \bibinfo{person}{Roberto
  Casado-Vara}.} \bibinfo{year}{2019}\natexlab{}.
\newblock \showarticletitle{A review of edge computing reference architectures
  and a new global edge proposal}.
\newblock \bibinfo{journal}{\emph{Future Generation Computer Systems}}
  \bibinfo{volume}{99} (\bibinfo{year}{2019}), \bibinfo{pages}{278 -- 294}.
\newblock
\showISSN{0167-739X}
\urldef\tempurl%
\url{https://doi.org/10.1016/j.future.2019.04.016}
\showDOI{\tempurl}


\bibitem[\protect\citeauthoryear{Stankovic and Ramamritham}{Stankovic and
  Ramamritham}{1990}]%
        {Stankovic}
\bibfield{author}{\bibinfo{person}{John Stankovic} {and}
  \bibinfo{person}{Krithi Ramamritham}.} \bibinfo{year}{1990}\natexlab{}.
\newblock \showarticletitle{What is predictability for real-time systems?}
\newblock \bibinfo{journal}{\emph{Real-Time Systems}} (\bibinfo{year}{1990}).
\newblock
\urldef\tempurl%
\url{https://doi.org/10.1007/BF01995673}
\showDOI{\tempurl}


\bibitem[\protect\citeauthoryear{{Taleb}, {Samdanis}, {Mada}, {Flinck},
  {Dutta}, and {Sabella}}{{Taleb} et~al\mbox{.}}{2017}]%
        {Taleb-7931566}
\bibfield{author}{\bibinfo{person}{T. {Taleb}}, \bibinfo{person}{K.
  {Samdanis}}, \bibinfo{person}{B. {Mada}}, \bibinfo{person}{H. {Flinck}},
  \bibinfo{person}{S. {Dutta}}, {and} \bibinfo{person}{D. {Sabella}}.}
  \bibinfo{year}{2017}\natexlab{}.
\newblock \showarticletitle{On Multi-Access Edge Computing: A Survey of the
  Emerging 5G Network Edge Cloud Architecture and Orchestration}.
\newblock \bibinfo{journal}{\emph{IEEE Communications Surveys Tutorials}}
  \bibinfo{volume}{19}, \bibinfo{number}{3} (\bibinfo{year}{2017}),
  \bibinfo{pages}{1657--1681}.
\newblock
\urldef\tempurl%
\url{https://doi.org/10.1109/COMST.2017.2705720}
\showDOI{\tempurl}


\bibitem[\protect\citeauthoryear{Tan, Zhu, Ge, and Xiong}{Tan
  et~al\mbox{.}}{2015}]%
        {tan2015utility}
\bibfield{author}{\bibinfo{person}{Liansheng Tan}, \bibinfo{person}{Zhongxun
  Zhu}, \bibinfo{person}{Fei Ge}, {and} \bibinfo{person}{Naixue Xiong}.}
  \bibinfo{year}{2015}\natexlab{}.
\newblock \showarticletitle{Utility maximization resource allocation in
  wireless networks: Methods and algorithms}.
\newblock \bibinfo{journal}{\emph{IEEE Transactions on systems, man, and
  cybernetics: systems}} \bibinfo{volume}{45}, \bibinfo{number}{7}
  (\bibinfo{year}{2015}), \bibinfo{pages}{1018--1034}.
\newblock


\bibitem[\protect\citeauthoryear{Tange, De~Donno, Fafoutis, and Dragoni}{Tange
  et~al\mbox{.}}{2019}]%
        {10.1145/3313150.3313228}
\bibfield{author}{\bibinfo{person}{Koen Tange}, \bibinfo{person}{Michele
  De~Donno}, \bibinfo{person}{Xenofon Fafoutis}, {and} \bibinfo{person}{Nicola
  Dragoni}.} \bibinfo{year}{2019}\natexlab{}.
\newblock \showarticletitle{Towards a Systematic Survey of Industrial IoT
  Security Requirements: Research Method and Quantitative Analysis}. In
  \bibinfo{booktitle}{\emph{Proceedings of the Workshop on Fog Computing and
  the IoT}} (Montreal, Quebec, Canada) \emph{(\bibinfo{series}{IoT-Fog
  ’19})}. \bibinfo{publisher}{Association for Computing Machinery},
  \bibinfo{address}{New York, NY, USA}, \bibinfo{pages}{56–63}.
\newblock
\showISBNx{9781450366984}
\urldef\tempurl%
\url{https://doi.org/10.1145/3313150.3313228}
\showDOI{\tempurl}


\bibitem[\protect\citeauthoryear{Thompson and Reimann}{Thompson and
  Reimann}{2018}]%
        {Platforms4CPS2018}
\bibfield{author}{\bibinfo{person}{Haydn Thompson} {and} \bibinfo{person}{Meike
  Reimann}.} \bibinfo{year}{2018}\natexlab{}.
\newblock \bibinfo{title}{Platforms4CPS Key Outcomes and Recommendations}.
\newblock \bibinfo{howpublished}{\url{https://www.platforms4cps.eu}}.
\newblock
\showISBNx{978-3-95663-184-9}


\bibitem[\protect\citeauthoryear{Tocz{\'e} and {Nadjm-Tehrani}}{Tocz{\'e} and
  {Nadjm-Tehrani}}{2018}]%
        {tocze2018}
\bibfield{author}{\bibinfo{person}{Klervie Tocz{\'e}} {and}
  \bibinfo{person}{Simin {Nadjm-Tehrani}}.} \bibinfo{year}{2018}\natexlab{}.
\newblock \showarticletitle{A {{Taxonomy}} for {{Management}} and
  {{Optimization}} of {{Multiple Resources}} in {{Edge Computing}}}.
\newblock \bibinfo{journal}{\emph{Wireless Communications and Mobile
  Computing}}  \bibinfo{volume}{2018} (\bibinfo{date}{June}
  \bibinfo{year}{2018}), \bibinfo{pages}{7476201}.
\newblock
\showISSN{1530-8669}
\urldef\tempurl%
\url{https://doi.org/10.1155/2018/7476201}
\showDOI{\tempurl}


\bibitem[\protect\citeauthoryear{T\"{o}rngren}{T\"{o}rngren}{1998}]%
        {TorngrenJRTS}
\bibfield{author}{\bibinfo{person}{Martin T\"{o}rngren}.}
  \bibinfo{year}{1998}\natexlab{}.
\newblock \showarticletitle{Fundamentals of Implementing Real-Time Control
  Applications in Distributed Computer Systems}.
\newblock \bibinfo{journal}{\emph{Real-Time Systems}}  \bibinfo{volume}{14}
  (\bibinfo{year}{1998}), \bibinfo{pages}{219 -- 250}.
\newblock
\urldef\tempurl%
\url{https://doi.org/10.1023/A:1007964222989}
\showDOI{\tempurl}


\bibitem[\protect\citeauthoryear{T\"{o}rngren, Asplund, Bensalem, McDermid,
  Passerone, Pfeifer, Sangiovanni-Vincentelli, and Sch\"{a}tz}{T\"{o}rngren
  et~al\mbox{.}}{2016}]%
        {toerngren16}
\bibfield{author}{\bibinfo{person}{Martin T\"{o}rngren},
  \bibinfo{person}{Fredrik Asplund}, \bibinfo{person}{Saddek Bensalem},
  \bibinfo{person}{John McDermid}, \bibinfo{person}{Roberto Passerone},
  \bibinfo{person}{Holger Pfeifer}, \bibinfo{person}{Alberto
  Sangiovanni-Vincentelli}, {and} \bibinfo{person}{Bernhard Sch\"{a}tz}.}
  \bibinfo{year}{2016}\natexlab{}.
\newblock \bibinfo{booktitle}{\emph{Characterization, analysis and
  recommendations for exploiting the opportunities of Cyber-Physical Systems}}.
\newblock \bibinfo{publisher}{Academic Press}, \bibinfo{pages}{3--14}.
\newblock
\showISBNx{978-0-128-03801-7}
\urldef\tempurl%
\url{https://doi.org/10.1016/B978-0-12-803801-7.00001-8}
\showDOI{\tempurl}


\bibitem[\protect\citeauthoryear{Törngren and Grogan}{Törngren and
  Grogan}{2018}]%
        {Torngren2018}
\bibfield{author}{\bibinfo{person}{Martin Törngren} {and}
  \bibinfo{person}{Paul Grogan}.} \bibinfo{year}{2018}\natexlab{}.
\newblock \showarticletitle{How to Deal with the Complexity of Future
  Cyber-Physical Systems?}
\newblock \bibinfo{journal}{\emph{Designs}} \bibinfo{volume}{2},
  \bibinfo{number}{4} (\bibinfo{date}{10} \bibinfo{year}{2018}),
  \bibinfo{pages}{40}.
\newblock
\showISSN{2411-9660}
\urldef\tempurl%
\url{https://doi.org/10.3390/designs2040040}
\showDOI{\tempurl}


\bibitem[\protect\citeauthoryear{(UL)}{(UL)}{2020}]%
        {UL4600}
\bibfield{author}{\bibinfo{person}{Underwriters~Laboratories (UL)}.}
  \bibinfo{year}{2020}\natexlab{}.
\newblock \bibinfo{title}{{ANSI/UL4600 - Evaluation of Autonomous Products}}.
\newblock
\newblock
\urldef\tempurl%
\url{https://ul.org/UL4600}
\showURL{%
\tempurl}


\bibitem[\protect\citeauthoryear{Wang, Sheng, Wang, Wang, and Li}{Wang
  et~al\mbox{.}}{2016}]%
        {wang2016mobile}
\bibfield{author}{\bibinfo{person}{Yanting Wang}, \bibinfo{person}{Min Sheng},
  \bibinfo{person}{Xijun Wang}, \bibinfo{person}{Liang Wang}, {and}
  \bibinfo{person}{Jiandong Li}.} \bibinfo{year}{2016}\natexlab{}.
\newblock \showarticletitle{Mobile-edge computing: Partial computation
  offloading using dynamic voltage scaling}.
\newblock \bibinfo{journal}{\emph{IEEE Transactions on Communications}}
  \bibinfo{volume}{64}, \bibinfo{number}{10} (\bibinfo{year}{2016}),
  \bibinfo{pages}{4268--4282}.
\newblock


\bibitem[\protect\citeauthoryear{Wieringa, Maiden, Mead, and Rolland}{Wieringa
  et~al\mbox{.}}{2005}]%
        {Wieringa}
\bibfield{author}{\bibinfo{person}{Roel Wieringa}, \bibinfo{person}{Neil
  Maiden}, \bibinfo{person}{Nancy Mead}, {and} \bibinfo{person}{Colette
  Rolland}.} \bibinfo{year}{2005}\natexlab{}.
\newblock \showarticletitle{Requirements Engineering Paper Classification and
  Evaluation Criteria: A Proposal and a Discussion}.
\newblock \bibinfo{journal}{\emph{Requir. Eng.}} \bibinfo{volume}{11},
  \bibinfo{number}{1} (\bibinfo{date}{Dec.} \bibinfo{year}{2005}),
  \bibinfo{pages}{102–107}.
\newblock
\showISSN{0947-3602}


\bibitem[\protect\citeauthoryear{Xu, Chen, and Ren}{Xu et~al\mbox{.}}{2017}]%
        {xu2017online}
\bibfield{author}{\bibinfo{person}{Jie Xu}, \bibinfo{person}{Lixing Chen},
  {and} \bibinfo{person}{Shaolei Ren}.} \bibinfo{year}{2017}\natexlab{}.
\newblock \showarticletitle{Online learning for offloading and autoscaling in
  energy harvesting mobile edge computing}.
\newblock \bibinfo{journal}{\emph{IEEE Transactions on Cognitive Communications
  and Networking}} \bibinfo{volume}{3}, \bibinfo{number}{3}
  (\bibinfo{year}{2017}), \bibinfo{pages}{361--373}.
\newblock


\bibitem[\protect\citeauthoryear{Yousefpour, Fung, Nguyen, Kadiyala, Jalali,
  Niakanlahiji, Kong, and Jue}{Yousefpour et~al\mbox{.}}{2019}]%
        {Yousefpour2019AllOneNeedsKnowFog}
\bibfield{author}{\bibinfo{person}{Ashkan Yousefpour}, \bibinfo{person}{Caleb
  Fung}, \bibinfo{person}{Tam Nguyen}, \bibinfo{person}{Krishna Kadiyala},
  \bibinfo{person}{Fatemeh Jalali}, \bibinfo{person}{Amirreza Niakanlahiji},
  \bibinfo{person}{Jian Kong}, {and} \bibinfo{person}{Jason~P. Jue}.}
  \bibinfo{year}{2019}\natexlab{}.
\newblock \showarticletitle{All {{One Needs}} to {{Know}} about {{Fog
  Computing}} and {{Related Edge Computing Paradigms}}: {{A Complete Survey}}}.
\newblock \bibinfo{journal}{\emph{Journal of Systems Architecture}}
  \bibinfo{volume}{98} (\bibinfo{date}{Sept.} \bibinfo{year}{2019}),
  \bibinfo{pages}{289--330}.
\newblock
\showISSN{13837621}
\urldef\tempurl%
\url{https://doi.org/10.1016/j.sysarc.2019.02.009}
\showDOI{\tempurl}


\bibitem[\protect\citeauthoryear{Yu, Wang, and Guo}{Yu et~al\mbox{.}}{2018}]%
        {yu2018energy}
\bibfield{author}{\bibinfo{person}{Hongyan Yu}, \bibinfo{person}{Quyuan Wang},
  {and} \bibinfo{person}{Songtao Guo}.} \bibinfo{year}{2018}\natexlab{}.
\newblock \showarticletitle{Energy-efficient task offloading and resource
  scheduling for mobile edge computing}. In \bibinfo{booktitle}{\emph{2018 IEEE
  international conference on networking, architecture and storage (NAS)}}.
  IEEE, \bibinfo{pages}{1--4}.
\newblock


\end{thebibliography}










\end{document}